\documentclass[pdflatex,sn-mathphys-num]{sn-jnl}

\usepackage{graphicx}%
\usepackage{multirow}%
\usepackage{amsmath,amssymb,amsfonts}%
\usepackage{amsthm}%
\usepackage{mathrsfs}%
\usepackage[title]{appendix}%
\usepackage{xcolor}%
\usepackage{textcomp}%
\usepackage{manyfoot}%
\usepackage{booktabs}%
\usepackage{algorithm}%
\usepackage{algorithmicx}%
\usepackage{algpseudocode}%
\usepackage{listings}%
\usepackage[table]{xcolor}
\usepackage{graphicx}
\newcommand{\E}{\mathbb{E}}

\newcommand{\norm}[1]{\left\lVert #1 \right\rVert}
\newcommand{\inner}[2]{\left\langle #1,#2 \right\rangle}
\usepackage{comment}
\usepackage{subcaption}
\usepackage{amsmath,amssymb,amsthm}
\usepackage{thmtools}   
\usepackage{braket}     

\newtheorem{assumption}{Assumption}
\DeclareMathOperator{\Tr}{Tr}
\newcommand{\X}{\mathcal{X}}
\newcommand{\Hil}{\mathcal{H}}
\newcommand{\W}{\mathcal{W}}
\newcommand{\A}{\mathcal{A}}
\newcommand{\Ss}{\mathcal{S}}
\newcommand{\Y}{\mathcal{Y}}
\newcommand{\Rad}{\mathfrak{R}}
\newcommand{\MMD}{\mathrm{MMD}}
\newcommand{\NAMEA}{\textsc{GENIE}}

\usepackage{xcolor}
\newcommand{\tbd}[1]{\textcolor{red}{[#1]}}

\begin{document}

\title[GENIE]{Noise-regularized training of quantum circuit Born machines under a finite shot budget}


\author[1]{\fnm{Hoang M.} \sur{Ngo}}\email{hoang.ngo@ufl.edu}

\author[1]{\fnm{Quan} \sur{Nguyen}}\email{quan.nguyen1@ufl.edu}

\author[2]{\fnm{Jung Taek} \sur{Seo}}\email{seojt@gachon.ac.kr}

\author[3]{\fnm{Wanli} \sur{Xing}}\email{wxx151@miami.edu}

\author*[1]{\fnm{My T.} \sur{Thai}}\email{mythai@cise.ufl.edu}

\affil*[1]{\orgdiv{Department of Computer and Information Science and Engineering},
\orgname{University of Florida},
\orgaddress{\street{1889 Museum Road},
\city{Gainesville}, \postcode{32611},
\state{Florida}, \country{USA}}}

\affil[2]{\orgdiv{Department of Smart Security},
\orgname{Gachon University},
\orgaddress{\street{1342 Seongnam-daero, Sujeong-gu},
\city{Seongnam-si}, \postcode{13120},
\state{Gyeonggi-do}, \country{Republic of Korea}}}

\affil[3]{\orgdiv{Department of Teaching and Learning, School of Education and Human Development},
\orgname{University of Miami},
\orgaddress{\street{5202 University Drive},
\city{Coral Gables}, \postcode{33124},
\state{Florida}, \country{USA}}}


\abstract{Quantum circuit Born machines (QCBMs) can learn distributions that are hard to sample from classically, but their value depends on generalizing from finite training data. Here, we study how hardware noise affects this generalization. We show that, for any noise process, the generalization gap of a QCBM trained with the maximum mean discrepancy is bounded by the complexity of the distributions the noisy model can reach. Scrambled local Pauli noise contracts this complexity but biases the model towards the uniform distribution, so residual noise acts as a regularizer whose optimal strength can be intermediate. We propose \NAMEA{}, a framework that realizes a chosen noise level with capped error mitigation and selects it during training from held-out data within a fixed execution budget, with provably bounded regret. In simulations on synthetic and census data, \NAMEA{} stays close to the best of four noise-control strategies, whereas every fixed noise level falls short in many settings.}




\maketitle


\section{Introduction}

Generative modelling, the task of learning an unknown probability distribution from examples and producing new samples from it, is one of the most natural tasks for a quantum computer. The reason is measuring a quantum state already yields a sample from a probability distribution~\cite{perdomo2018opportunities}. Parameterized quantum circuits can prepare output distributions that are believed to be hard to sample from classically~\cite{bremner2016average,arute2019quantum,coyle2020born}, and quantum generative models admit provable expressivity and learnability separations from classical models~\cite{gao2018quantum,gao2022enhancing,sweke2021quantum}. The most direct realization of this idea is the quantum circuit Born machine (QCBM)~\cite{benedetti2019generative,liu2018differentiable}, whose model distribution is given by the Born rule, $q_{\boldsymbol{\theta}}(\boldsymbol{x})=|\langle\boldsymbol{x}|U(\boldsymbol{\theta})|\boldsymbol{0}\rangle|^{2}$, for a parameterized circuit $U(\boldsymbol{\theta})$. Every circuit execution thus returns an exact sample, and because only samples rather than likelihoods are accessible, QCBMs are trained with sample-based losses such as the maximum mean discrepancy (MMD)~\cite{gretton2012kernel,liu2018differentiable,rudolph2024trainability}. QCBMs have been demonstrated on trapped-ion hardware~\cite{rudolph2022generation}, high-energy physics~\cite{kiss2022conditional}, combinatorial optimization~\cite{alcazar2024enhancing} and drug discovery~\cite{vakili2025quantum}.

To be practically useful, a Born machine must generalize beyond its training set rather than reproduce it. For a generative model, \emph{generalization} is defined as the closeness of its output distribution to the unknown target distribution $p$, rather than to the empirical distribution of the $m$ training samples. Operationally, it is the ability to produce valid samples that never appeared in training~\cite{gili2024generalization}. For a sufficiently expressive model, driving the training loss to zero instead reproduces the empirical distribution, which places all of its mass on the observed bitstrings and none on valid but unseen ones~\cite{rudolph2024trainability}. This is the generative form of overfitting, and Born machines are particularly exposed to it. Specifically, their output space contains $2^{n}$ bitstrings while practical training sets are only polynomially large, so the empirical distribution leaves almost every valid bitstring unobserved.
Generalization bounds and regularizers developed for supervised quantum learning~\cite{abbas2021power,banchi2021generalization,caro2022generalization,du2022efficient,kobayashi2022overfitting} do not transfer directly, because they act on a predictor that maps inputs to labels, whereas a Born machine has neither and outputs only the distribution it samples. Recent studies have begun to examine generalization in Born machines by bounding the excess MMD risk of an idealized empirical minimizer~\cite{du2022theory}, benchmarking the validity, novelty and coverage of generated samples~\cite{gili2023generalize,gili2024generalization}, and showing that a converged MMD is an unreliable indicator of generalization~\cite{raj2026train}.
However, these works evaluate generalization rather than control it, and they offer no mechanism for regularizing the finite-shot, noisy training that Born machines undergo on real hardware.

In classical learning, injecting stochasticity into training is a common way to improve generalization, whether through input noise~\cite{bishop1995training}, dropout~\cite{srivastava2014dropout} or stochastic gradient descent~\cite{hardt2016train}. Quantum hardware provides such stochasticity intrinsically from gate errors and decoherence. In supervised quantum learning, analyses~\cite{bu2023effects,du2022efficient,khanal2025data} show that it can act as an implicit regularizer~\cite{heyraud2022noisy}, with a finite optimal noise level that can be targeted by tuning the noise~\cite{zhang2026theory,somogyi2025method,scala2026noise}, although excessive noise destroys trainability~\cite{wang2021noise}.
For Born machines, exploiting this noise is more challenging.
The noisy output distribution is itself the model, so the noise that suppresses overfitting also distorts what is learned. 
Moreover, the noise level is fixed by the hardware, and lowering it requires error mitigation whose sampling cost grows exponentially with the amount of noise removed~\cite{temme2017error,mari2021extending,takagi2022fundamental,quek2024exponentially}.
To build a noise-based regularizer for QCBMs, we must therefore answer two questions:
(i) how does residual hardware noise affect the generalization of an MMD-trained Born machine?
(ii) how can the amount of residual noise be selected and realized during training within a finite budget of circuit executions?

\begin{figure}[t]
    \centering
\includegraphics[width=\linewidth]{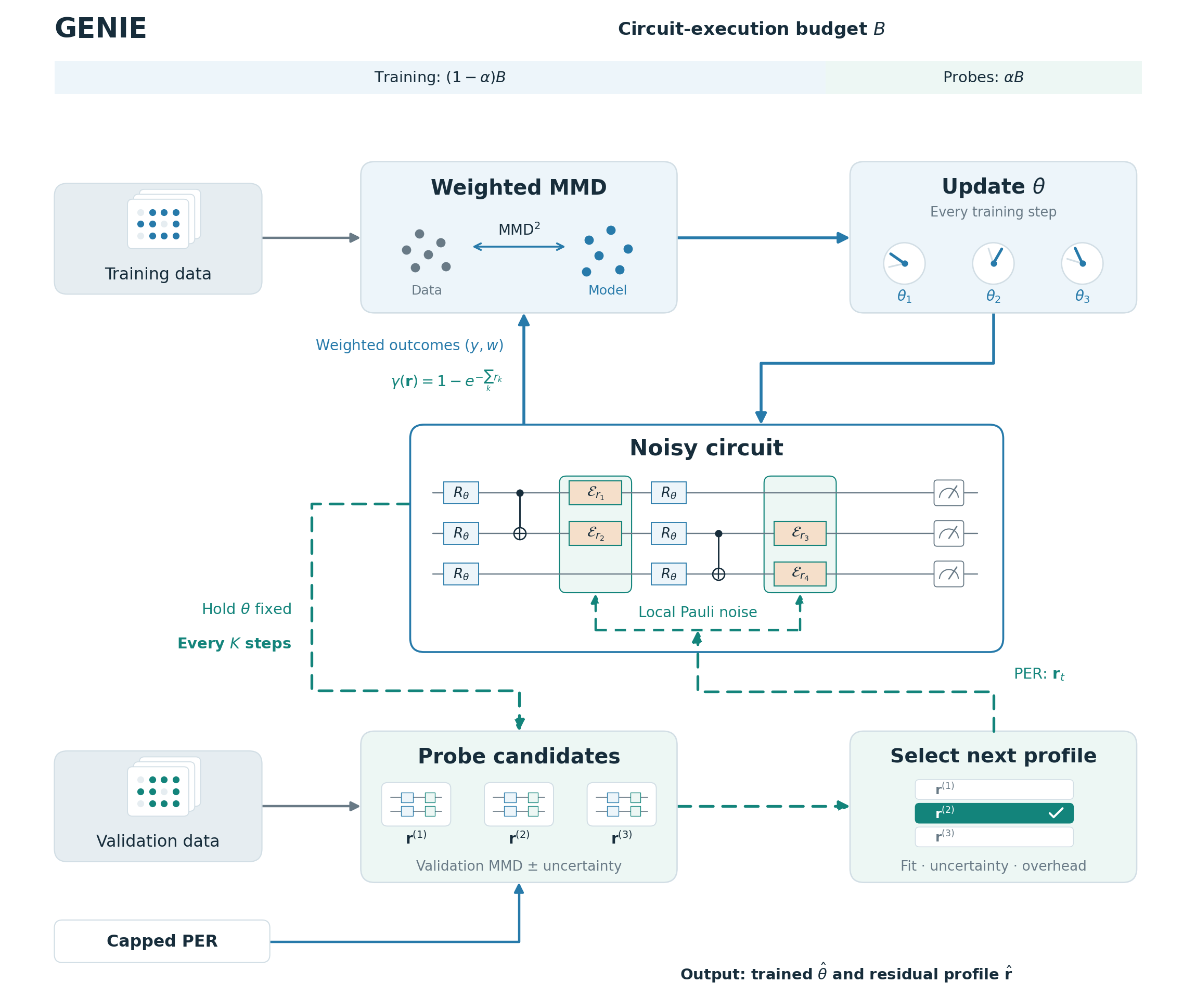}
    \caption{\textbf{Overview of GENIE.}
    The circuit-execution budget $B$ is divided between training,
    $(1-\alpha)B$, and probing, $\alpha B$.
    Circuit parameters $\theta$ are updated using the MMD loss estimated
    from training data and outcomes weighted by probabilistic error reduction (PER) $(y,w)$, where $y$ is a
    measured bitstring and $w$ is its signed PER weight.
    Every $K$ steps, $\theta$ is held fixed while candidate residual-noise
    profiles from capped PER allocation are evaluated on held-out
    validation data. The next profile is selected using validation fit,
    estimation uncertainty, and sampling overhead.
    The selected profile specifies target local Pauli-noise rates $r_k$
    throughout the circuit, with $\gamma(\mathbf{r})$ summarizing the
    effective global noise under the assumed model.
    \NAMEA{} returns the trained parameters $\hat{\theta}$ and the selected
    residual-noise profile $\hat{\mathbf{r}}$.}
    \label{fig:genie-overview}
\end{figure}

In this work, we answer these questions and develop GENIE (generalization with inherent noise), a framework that treats residual hardware noise as a tunable regularizer for MMD-trained Born machines (Fig.~\ref{fig:genie-overview}).
Our contributions are as follows.
\begin{itemize}
    \item \textbf{Generalization theory.} We derive a bound on the empirical-to-population MMD gap of QCBMs that holds for any noise process and training algorithm. In addition, we specialize it to scrambled local Pauli noise, where residual noise contracts the gap by the retained-signal factor $1-\gamma$. The bound makes the regularization effect of residual noise explicit, while showing that the same noise biases the model towards the uniform distribution, so the amount of noise to retain must be chosen.

\item \textbf{Adaptive noise regularization.}
We propose \NAMEA{}, a framework that adaptively selects the residual noise for MMD-trained QCBMs under a fixed circuit-execution budget. In general, first it realizes each level with capped probabilistic error reduction, which removes just enough hardware noise while keeping the remaining noise close to white noise. Then, it periodically spends a small portion of the budget to probe candidate noise levels on held-out data, and then selects the level for the next training block based on validation performance and the budget remaining for training.
We establish a cumulative selection regret bound that quantifies how limited validation data, measurement precision, and the frequency of noise adjustments affect the controller's decisions along the training trajectory.

    \item \textbf{Empirical validation.} We compare \NAMEA{} with training at three fixed noise levels, namely no, half and full mitigation, under matched budgets of circuit executions. In simulations on three synthetic datasets with $12$ qubits, it is within $1\%$ of the best of the four methods in $64\%$ of $144$ conditions and has a median regret of $0.3\%$, whereas the fixed levels have median regrets between $8.4\%$ and $22.7\%$. On the Adult census dataset with $10$ qubits at strong noise, it gives the most accurate population shares, income rates within groups and income predictions of the four methods.
\end{itemize}


\section{Results}

\subsection{Preliminaries}

This subsection introduces the model, the noise and the loss, and the two quantities in which all of our results are stated, which are, the witness function and the Rademacher complexity.

\paragraph{Born machines under hardware noise.}
Let $\X=\{0,1\}^n$ be the set of $n$-bit strings. A Born machine applies a parameterized circuit $U(\theta)$, $\theta\in\Theta$, to a fixed input state $\rho_0$, typically $\ket{0}^{\otimes n}$, and measures all qubits in the computational basis. By the Born rule, its ideal output distribution is
\begin{equation}
q_\theta^{0}(x)=\bra{x}U(\theta)\rho_0U(\theta)^\dagger\ket{x},\qquad x\in\X .
\label{eq:prelim-born}
\end{equation}
On hardware, the circuit runs under a noise process $\mathcal N$, which may include gate errors, correlated errors and readout errors. If $\rho_\theta^{\mathcal N}$ denotes the resulting output state, each circuit execution returns one sample from the noisy distribution
\begin{equation}
q_\theta^{\mathcal N}(x)=\bra{x}\rho_\theta^{\mathcal N}\ket{x}.
\label{eq:prelim-noisy-born}
\end{equation}
The probabilities $q_\theta^{\mathcal N}(x)$ are never observed, only samples from them. Therefore, the Born machine is an implicit generative model, and the distribution that is trained and evaluated is the noisy $q_\theta^{\mathcal N}$, not the ideal $q_\theta^0$.

To model the noise, we use the standard model of local gate noise. Each of the $s$ two-qubit gates in the circuit is followed by an independent stochastic Pauli channel on each of the two qubits it acts on. Indexing these $L=2s$ noisy locations by $k$, the channel at location $k$ is
\begin{equation}
\mathcal E_k(\rho)=(1-\epsilon_k)\rho+\epsilon_k\sum_{P\in\{X,Y,Z\}}\pi_k(P)\,P\rho P, \qquad\sum_P\pi_k(P)=1,
\label{eq:characterized-pauli-channel}
\end{equation}
with probability $\epsilon_k$ an error occurs, and the error is a Pauli $P$ drawn from the distribution $\pi_k$. Randomized compiling brings general gate noise into this form, after which the error rates can be measured~\cite{wallman2016noise,vandenberg2023probabilistic}. We call $\boldsymbol\epsilon=(\epsilon_1,\ldots,\epsilon_L)$ the hardware noise profile. It is fixed by the device, and later we reduce it selectively to a residual profile $\mathbf r$ with $0\le r_k\le\epsilon_k$.

\paragraph{Learning problem and MMD loss.}
The target distribution $p$ on $\X$ is unknown and is accessible only through a training sample $\Ss=(x_1,\ldots,x_m)$ of $m$ independent draws from $p$. The goal is a model whose distribution is close to $p$ itself, not merely to the $m$ observed strings. Because only samples of the model are available, we measure closeness with a kernel $\kappa:\X\times\X\to[0,1]$ between bit strings that is positive semidefinite and satisfies $\kappa(x,x)=1$. A standard choice for Born machines is the Gaussian kernel on the Hamming distance $d_H$, $\kappa(x,y)=\frac1c\sum_{i=1}^c\exp[-d_H(x,y)/(2\tau_i^2)]$, with bandwidths $\tau_1,\ldots,\tau_c$~\cite{liu2018differentiable,rudolph2024trainability}. The kernel defines a reproducing kernel Hilbert space (RKHS) $\Hil$ of functions on $\X$, in which $\kappa(x,\cdot)$ serves as a feature vector of the string $x$, with $\inner{\kappa(x,\cdot)}{\kappa(y,\cdot)}_{\Hil}=\kappa(x,y)$. Averaging these features over a distribution $\nu$ gives its kernel mean embedding, and the distance between two embeddings is the maximum mean discrepancy (MMD)~\cite{gretton2012kernel}:
\begin{equation}
\mu_\nu:=\mathbb E_{x\sim\nu}\bigl[\kappa(x,\cdot)\bigr]\in\Hil,
\qquad
\MMD_\kappa(\nu,\nu'):=\norm{\mu_\nu-\mu_{\nu'}}_{\Hil}.
\label{eq:prelim-mmd}
\end{equation}
The population risk of the noisy model is its squared MMD to the target,
\begin{equation}
R_{\mathcal N}(\theta):=\MMD_\kappa^2\bigl(q_\theta^{\mathcal N},p\bigr)
=\underbrace{\mathbb E_{y,y'\sim q_\theta^{\mathcal N}}\bigl[\kappa(y,y')\bigr]}_{\text{model--model}}
+\underbrace{\mathbb E_{x,x'\sim p}\bigl[\kappa(x,x')\bigr]}_{\text{target--target}}
-2\,\underbrace{\mathbb E_{x\sim p,\,y\sim q_\theta^{\mathcal N}}\bigl[\kappa(x,y)\bigr]}_{\text{coupling}},
\label{eq:prelim-risk}
\end{equation}
where all draws are independent. The coupling term rewards model samples that resemble the data, the model--model term penalizes a model whose samples all resemble each other, and the target--target term is a constant.

\paragraph{Training, evaluation and the generalization gap.}
A learning algorithm $\A_{\mathcal N}$ maps the training sample to parameters $\hat\theta_{\mathcal N}=\A_{\mathcal N}(\Ss)$. It may be randomized, for example through the initialization, stochastic optimization and the finite number of shots used in training. Because $p$ is unknown, the risk of the learned model can only be estimated. To do so, we draw $N$ fresh samples $\Y=(y_1,\ldots,y_N)$ from the learned circuit, independent of $\Ss$ and of the training randomness given $\hat\theta_{\mathcal N}$, and replace each expectation in equation~\eqref{eq:prelim-risk} by an average over distinct pairs:
\begin{align}
\label{eq:prelim-ustat}
\widehat R^{\mathcal N}_{\Ss,\Y}(\theta)
:=\frac{1}{m(m-1)}\sum_{i\neq i'}\kappa(x_i,x_{i'})
+\frac{1}{N(N-1)}\sum_{j\neq j'}\kappa(y_j,y_{j'})
&-\frac{2}{mN}\sum_{i=1}^m\sum_{j=1}^N\kappa(x_i,y_j),\\
&\qquad m,N\ge2\nonumber.
\end{align}
For a fixed $\theta$, this U-statistic is an unbiased estimate of $R_{\mathcal N}(\theta)$. For the learned parameters it is not, because $\hat\theta_{\mathcal N}$ was chosen using the same training strings that appear in the estimate. The generalization gap measures the resulting optimism,
\begin{equation}
\Delta_{\mathcal N}(m):=\mathbb E_{\Ss,\A_{\mathcal N},\Y}\Bigl[R_{\mathcal N}(\hat\theta_{\mathcal N})-\widehat R^{\mathcal N}_{\Ss,\Y}(\hat\theta_{\mathcal N})\Bigr].
\label{eq:prelim-gap}
\end{equation}
Therefore, a low empirical risk certifies a low population risk only if the gap is small. For example, a model that simply reproduces the empirical distribution $\hat p_m$ of the training strings has an empirical risk of at most zero on average over $\Y$, although its population risk $\MMD^2_\kappa(\hat p_m,p)$ is positive whenever $\hat p_m\neq p$. 

\paragraph{Witness functions.}
Only the coupling term in equation~\eqref{eq:prelim-risk} involves both the model and the data. It is described by the witness function of a generator,
\begin{equation}
h_\theta^{\mathcal N}(x):=\inner{\mu_{q_\theta^{\mathcal N}}}{\kappa(x,\cdot)}_{\Hil}
=\mathbb E_{y\sim q_\theta^{\mathcal N}}\bigl[\kappa(x,y)\bigr]\in[0,1],
\label{eq:prelim-witness}
\end{equation}
which is large at strings $x$ that resemble what the model generates and small elsewhere. Averaging the witness over the target recovers the coupling term. For a function $f$ on $\X$, we write
\begin{equation}
P_mf:=\frac1m\sum_{i=1}^mf(x_i),
\qquad
Pf:=\mathbb E_{x\sim p}\bigl[f(x)\bigr]
\label{eq:prelim-averages}
\end{equation}
for its average over the training sample and over the target. In this language, a learned model overfits when its witness is larger on the training strings than on typical target strings, $(P_m-P)h^{\mathcal N}_{\hat\theta_{\mathcal N}}>0$. Adding a constant to a function does not change this difference,
\begin{equation}
(P_m-P)(f+c)=(P_m-P)f,\qquad c\in\mathbb R,
\label{eq:prelim-centering}
\end{equation}
so constant parts of a witness never contribute to overfitting. The witness class
\begin{equation}
\W^{\mathcal N}:=\bigl\{h_\theta^{\mathcal N}:\theta\in\Theta\bigr\},
\qquad
\W_\pm^{\mathcal N}:=\W^{\mathcal N}\cup\bigl(-\W^{\mathcal N}\bigr),
\label{eq:prelim-witness-class}
\end{equation}
contains one witness for each distribution that the noisy device can reach. It depends on the noise and on which distributions are reachable, but not on how many parameters are used to reach them.

\paragraph{Rademacher complexity.}
How much a learned witness can overfit depends on how rich the witness class is. We measure richness with the Rademacher complexity~\cite{bartlett2002rademacher},
\begin{equation}
\Rad(\mathcal F):=\mathbb E_{\Ss\sim p^{\otimes m}}\,\mathbb E_{\sigma}\Bigl[\sup_{f\in\mathcal F}\frac1m\sum_{i=1}^m\sigma_if(x_i)\Bigr],
\label{eq:prelim-rademacher}
\end{equation}
where $\sigma_1,\ldots,\sigma_m$ are independent uniform random signs. Each draw of the signs splits the training strings at random into two groups, and the supremum asks how well some function in $\mathcal F$ can be large on one group and small on the other. A rich class can follow any such random split, and hence the accidental features of a particular sample while a poor class cannot. Every witness has RKHS norm at most one, which implies $\Rad(\W^{\mathcal N}_\pm)\le1/\sqrt m$ for any noise process. Our results show how noise lowers the constant in front of $1/\sqrt m$.


\subsection{Residual noise contracts the generalization gap}

In this subsection, we bound the generalization gap, first for an arbitrary noise process and then for local Pauli noise, where the effect of noise becomes explicit. Throughout, we make one assumption on the kernel, which exposes the structure on which noise acts.

\begin{assumption}[Translation-invariant kernel]
\label{ass:translation-invariant-kernel}
The positive-semidefinite kernel
$\kappa:\X\times\X\rightarrow[0,1]$ satisfies
\begin{equation}
\kappa(x,y)
=
\phi(x\oplus y),
\qquad
\phi(\mathbf 0)=1,
\label{eq:ti-kernel}
\end{equation}
for some $\phi:\X\rightarrow[0,1]$.
\end{assumption}

This assumption covers the Hamming-distance kernels commonly used for MMD training of Born machines, including the Gaussian kernels above~\cite{liu2018differentiable,coyle2020born}. To bound the gap, note that the target--target term in equation~\eqref{eq:prelim-risk} does not depend on $\theta$ and the model--model term is estimated from samples drawn after training, so only the coupling term contributes (i.e., $\Delta_{\mathcal N}(m)=2\,\mathbb E[(P_m-P)h^{\mathcal N}_{\hat\theta_{\mathcal N}}]$), which does not depend on $N$ (Lemma~\ref{lem:expected-gap-identity}, Methods). Because the learned witness lies in $\W^{\mathcal N}$, its deviation is at most the largest deviation over the class, which symmetrization bounds by the Rademacher complexity. This gives our first result.

\begin{restatable}[Noise-agnostic generalization bound]{theorem}{theoremnoiseagnostic}
\label{thm:noise-agnostic-generalization}
Let $m \geq 2$ and $\A_{\mathcal N}$ be any possibly randomized learning
algorithm with the noise model $\mathcal N$, we have
\begin{equation}
\left|\Delta_{\mathcal N}(m)\right|
\le
4\,
\Rad\left(\W_\pm^{\mathcal N}\right).
\label{eq:noise-agnostic-generalization-bound}
\end{equation}
\end{restatable}

Theorem~\ref{thm:noise-agnostic-generalization} separates generalization from the details of the training algorithm and of the noise model. It holds for any noise process, including correlated and readout errors, and all dependence on the noise is captured by the reachable witness class. Therefore, noise improves the bound only when it reduces the complexity of that class. In the rest of this subsection, we quantify how much local Pauli noise reduces it.

\paragraph{Scrambled local Pauli noise.}
We now make this reduction explicit for local Pauli noise, a standard model of gate errors into which randomized compiling converts general gate noise~\cite{wallman2016noise}. Because noise acts on the output distribution, we first describe how a witness depends on that distribution. Bit strings form a group under bitwise addition, and the corresponding Fourier basis consists of the parities $\chi_A(x):=(-1)^{\sum_{j\in A}x_j}$ of subsets $A\subseteq[n]$ of qubits. We write
\begin{equation}
\widehat\phi(A):=2^{-n}\sum_{t\in\X}\phi(t)\chi_A(t),
\qquad
\widehat q(A):=\mathbb E_{x\sim q}\bigl[\chi_A(x)\bigr]
\label{eq:fourier-conventions}
\end{equation}
for the Fourier coefficients of the kernel and of a distribution $q$. The coefficient $\widehat q(A)$ is the parity moment of $q$ on the qubits in $A$, that is, the expectation value of the Pauli string $Z_A=\bigotimes_{j\in A}Z_j$, and $\widehat\phi(A)\ge0$ is the weight that the kernel places on it. Under Assumption~\ref{ass:translation-invariant-kernel}, a noiseless witness is then a kernel-weighted sum of these moments,
\begin{equation}
h_\theta^0(x)=\sum_{A\subseteq[n]}\widehat\phi(A)\,\widehat{q_\theta^0}(A)\,\chi_A(x).
\label{eq:witness-fourier}
\end{equation}
The term with $A=\varnothing$ equals the same constant $\widehat\phi(\varnothing)$ for every generator, so by equation~\eqref{eq:prelim-centering} it cannot cause overfitting. Therefore, overfitting can arise only from the nonconstant moments, and these are exactly what noise erodes.

Consider a circuit of $s$ two-qubit gates in which every gate is followed by local stochastic Pauli errors of rate $\epsilon$ on the qubits it acts on. Although each error is local, a sufficiently scrambling circuit spreads it over the entire output distribution. Dalzell \textit{et al.}~\cite{dalzell2024random} showed that, for random circuits under weak local noise, the output distribution is close to a mixture of the ideal and uniform distributions, with effective noise rate $\gamma=1-\exp(-2s\epsilon\pm\mathcal O(s\epsilon^2))$ and an approximation error of order $(1-\gamma)\epsilon\sqrt s$. Such white-noise models are also commonly used to describe the accumulated noise of hardware~\cite{vovrosh2021simple,urbanek2021mitigating}. We adopt this description for the circuits we train and measure its error in the geometry of the loss.

\begin{assumption}[Scrambled local Pauli noise]
\label{ass:effective-depolarizing}
Suppose that each circuit contains $s$ two-qubit gates, each followed by local Pauli errors of rate $\epsilon$ on both of its qubits. The circuit family scrambles these errors sufficiently that
the resulting output distribution is uniformly approximated by an effective
white-noise model. Specifically, there exist an effective noise rate
$\gamma\in[0,1]$ and a constant $c_{\mathrm{scr}}>0$ such that
\begin{align}
q_\theta^\gamma(x)
&:=
(1-\gamma)q_\theta^0(x)
+
\gamma u(x),
\qquad
u(x)=2^{-n},
\label{eq:effective-depolarizing-distribution}
\\
\sup_{\theta\in\Theta}
\operatorname{MMD}_\kappa
\left(
q_\theta^{\mathcal N},
q_\theta^\gamma
\right)
&\leq
c_{\mathrm{scr}}
(1-\gamma)\epsilon\sqrt{s}
=:
\varepsilon_{\mathrm{Pauli}}.
\label{eq:effective-depolarizing-assumption}
\end{align}
\end{assumption}

Since trained circuits are not random, this is a modelling assumption about the circuit family.
Because the uniform distribution has no nonconstant parity moments, white noise scales every nonconstant moment, and hence the part of every witness that can overfit, by $1-\gamma$. The part of the physical noise that is not white changes each witness by at most $\varepsilon_{\mathrm{Pauli}}$ in RKHS norm (Methods). Combining these two facts with Theorem~\ref{thm:noise-agnostic-generalization} gives our main generalization result.

\begin{restatable}[Generalization under scrambled local Pauli noise]{theorem}{theorempauli}
\label{thm:effective-depolarizing-generalization}
Suppose that Assumptions~\ref{ass:translation-invariant-kernel} and
\ref{ass:effective-depolarizing} hold and that, for some $C\geq1$,
\begin{equation}
\sup_{\theta\in\Theta}
2^n\sum_{x\in\X}q_\theta^0(x)^2
\leq
C.
\label{eq:effective-depolarizing-anticoncentration}
\end{equation}
Let $m,N\geq2$. Then the expected generalization gap satisfies
\begin{equation}
\left|\Delta_{\mathcal N}(m)\right|
\leq
4(1-\gamma)
\frac{\Lambda_C}{\sqrt m}
+
\frac{2\varepsilon_{\mathrm{Pauli}}}{\sqrt m},
\label{eq:effective-depolarizing-expected-bound}
\end{equation}
where $\Lambda_C:=\min\bigl\{\sqrt{(C-1)\sum_{A\neq\varnothing}\widehat\phi(A)^2},\,1-\widehat\phi(\varnothing)\bigr\}$.
\end{restatable}

Theorem~\ref{thm:effective-depolarizing-generalization} makes the effect of noise explicit. The constant $\Lambda_C$ is the complexity of the noiseless witness class. In its first branch, $C-1$ bounds the total squared magnitude of the nonconstant parity moments that the circuits can reach, which vanishes for the uniform distribution and grows as the outputs concentrate, and $\sum_{A\neq\varnothing}\widehat\phi(A)^2$ measures how strongly the loss weighs those moments. Because $C$ can be as large as $2^n$, for example when the circuit can output a single bit string, the second branch caps the complexity of every circuit family at a constant below one. Neither branch depends on the number of parameters, unlike supervised quantum learning bounds that grow with the number of trainable gates~\cite{caro2022generalization}. White noise contracts this complexity by the retained-signal factor $1-\gamma$. Specifically, uniform samples carry no information about which strings appeared in training, so retained noise dilutes the part of the model that can memorize them, as if a noiseless model were trained on $m/(1-\gamma)^2$ samples. Noise that is not white enters only through $\varepsilon_{\mathrm{Pauli}}$, which also decays as $m^{-1/2}$. Therefore, noise helps generalization to the extent that it acts as white noise, and this holds not only on average but for a single training run.

\begin{restatable}[High-probability generalization bound]{corollary}{corollarypauli}
\label{cor:effective-depolarizing-high-probability}
Under the conditions of
Theorem~\ref{thm:effective-depolarizing-generalization}, for every
$\beta\in(0,1)$, with probability at least $1-\beta$ over
$\Ss$, $\A_{\mathcal N}$, and $\Y$,
\begin{align}
\left| R_{\mathcal N}(\hat\theta_{\mathcal N}) - \widehat R_{\Ss,\Y}^{\mathcal N}(\hat\theta_{\mathcal N}) \right|
&\leq 4(1-\gamma) \frac{\Lambda_C}{\sqrt m}
+ 2(1-\gamma) \sqrt{\frac{2\log(4/\beta)}{m}} \notag\\
&\quad+ \frac{2\varepsilon_{\mathrm{Pauli}}}{\sqrt m} \left( 1+\sqrt{2\log(4/\beta)} \right)
+ \sqrt{ \frac{\log(8/\beta)} {2\lfloor m/2\rfloor} }
+ \sqrt{ \frac{8\log(8/\beta)}{N} }.
\label{eq:effective-depolarizing-high-probability}
\end{align}
\end{restatable}

Corollary~\ref{cor:effective-depolarizing-high-probability} adds two observations to Theorem~\ref{thm:effective-depolarizing-generalization}. First, the first three terms all carry the factor $1-\gamma$, the third through $\varepsilon_{\mathrm{Pauli}}$, so noise reduces not only the expected gap but also its fluctuation across draws of the training sample, and a single-run guarantee costs only a logarithmic factor in $1/\beta$. Second, the last two terms do not depend on $\gamma$. They come from estimating the target--target term with $m$ samples and the model-dependent terms with $N$ evaluation shots, and they form a floor of order $\sqrt{\log(1/\beta)/m}+\sqrt{\log(1/\beta)/N}$ that no amount of noise can lower. The shot term can be made negligible by measuring more, so the floor that remains is set by the training data. Because the contracted terms and this floor both decay as $m^{-1/2}$, their ratio does not depend on $m$. In terms of the bound, noise therefore yields a substantial reduction only while $(1-\gamma)\Lambda_C$ exceeds $\sqrt{\log(1/\beta)}$ up to constants, which favours expressive circuit families with concentrated outputs, precisely the models most prone to overfitting. Beyond that point, additional noise barely shrinks the gap but still pulls the model towards the uniform distribution, a cost we quantify in the next subsection.
 
\subsection{Realizing a target noise level by capped error reduction}

The previous subsection shows that retained noise contracts the bound on the generalization gap. Noise, however, also changes the model itself. Under the white-noise model, $\mu_{q^\gamma_\theta}=(1-\gamma)\mu_{q^0_\theta}+\gamma\mu_u$, so the population loss becomes
\begin{equation}
\MMD^2_\kappa\bigl(q^\gamma_\theta,p\bigr)=\norm{\bigl(\mu_{q^0_\theta}-\mu_p\bigr)-\gamma\bigl(\mu_{q^0_\theta}-\mu_u\bigr)}_{\Hil}^2 .
\label{eq:white-noise-bias}
\end{equation}
Noise thus pulls the kernel mean embedding of the model towards that of the uniform distribution, and even a perfect fit, $q^0_\theta=p$, incurs the loss $\gamma^2\MMD^2_\kappa(p,u)$. The effective noise level therefore acts as a regularization parameter that trades a smaller gap against the loss of the nonuniform structure the target requires, and its best value can be intermediate. 

Therefore, using noise as a regularizer raises two practical questions. The first is how to realize a chosen noise level. The hardware supplies a fixed amount of noise, so a lower level can only be reached by removing part of it with error mitigation. The second is which level to choose, because the best value is not known before training. In this work, we propose \NAMEA{}, a noise-based regularization framework that answers these two questions with two components (see Fig.~\ref{fig:genie-overview} for an overview). In this subsection, we present the first component, capped probabilistic error reduction (capped PER), which realizes a target noise level by removing just enough hardware noise while keeping the remaining noise close to white noise. 


\paragraph{Residual noise and its cost.}
We start from the characterized hardware profile $\boldsymbol\epsilon$ of equation~\eqref{eq:characterized-pauli-channel}. To lower the noise, we choose a residual profile $\mathbf r=(r_1,\ldots,r_L)$ with $0\le r_k\le\epsilon_k$, which keeps each $\pi_k$ but reduces its rate to $r_k$. We implement it with probabilistic error reduction (PER)~\cite{mari2021extending}, the partial variant of probabilistic error cancellation~\cite{temme2017error,vandenberg2023probabilistic}. PER inserts randomly sampled Pauli corrections and attaches to each measurement outcome a signed weight of magnitude $\Gamma(\mathbf r)$. Weighted averages are then unbiased for expectation values under the residual noise, with a standard error that grows as $\Gamma(\mathbf r)/\sqrt N$. In the weak-noise regime, this cost takes a simple form (Lemma~\ref{lem:profile-cost}, Methods),
\begin{equation}
\Gamma(\mathbf r)\simeq\exp\bigl(2\norm{\boldsymbol\epsilon-\mathbf r}_1\bigr).
\label{eq:profile-cost-main}
\end{equation}
The cost depends only on the total amount of noise removed, and it grows exponentially with it. The MMD is more expensive still, because its model--model term multiplies two weights, so a fixed precision can require about $\Gamma(\mathbf r)^4$ times more circuit executions than without mitigation. Every unit of noise that we keep therefore saves an exponential number of executions.

\paragraph{Effective noise of a residual profile.}
To connect a residual profile to Theorem~\ref{thm:effective-depolarizing-generalization}, we assume Assumption~\ref{ass:effective-depolarizing} in a profile-dependent form. Writing $q_\theta^{\mathbf r}$ for the output distribution under $\mathbf r$, its effective noise level and its distance from white noise are
\begin{align}
\gamma(\mathbf r)&:=1-\exp\bigl(-\norm{\mathbf r}_1\bigr),
\label{eq:gamma-from-residual-profile}\\
\sup_{\theta\in\Theta}\MMD_\kappa\bigl(q_\theta^{\mathbf r},q_\theta^{\gamma(\mathbf r)}\bigr)&\le c_{\mathrm{scr}}\exp\bigl(-\norm{\mathbf r}_1\bigr)\norm{\mathbf r}_2,
\label{eq:profile-dependent-scrambling}
\end{align}
which recovers the scaling of Assumption~\ref{ass:effective-depolarizing} for homogeneous rates up to a constant factor. Like the cost, the effective level depends only on the total retained noise $\norm{\mathbf r}_1$. The distance from white noise, in contrast, also depends on how the retained noise is spread across the circuit, through $\norm{\mathbf r}_2$. Where the noise is kept therefore matters only for how white it is.

\paragraph{Capped allocation problem.}
We can now state the design problem that capped PER solves. A target $\gamma_\star$ specifies the largest effective noise level we are willing to keep. Many residual profiles stay within this level, and they differ in two respects that matter, namely how much mitigation they require and how close their remaining noise is to white noise. We therefore seek the residual profile that meets the target at the lowest cost while keeping its distance from white noise below a tolerance $\varepsilon_{\mathrm{tol}}$,

\begin{align}
\begin{aligned}
\min_{\mathbf r\in\mathbb R^L}
\quad&
\exp\left(
2\norm{\boldsymbol\epsilon-\mathbf r}_1
\right)
\\
\text{subject to}\quad&
1-\exp\left(-\norm{\mathbf r}_1\right)
\leq
\gamma_\star,
\\
&
c_{\mathrm{scr}}
\exp\left(-\norm{\mathbf r}_1\right)
\norm{\mathbf r}_2
\leq
\varepsilon_{\mathrm{tol}},
\\
&
0\leq r_k\leq\epsilon_k,
\qquad
k=1,\ldots,L.
\end{aligned}
\label{eq:capped-per-design}
\end{align}
Because the cost and the level depend only on $R:=\norm{\mathbf r}_1$, the problem separates into deciding how much noise to keep and deciding where to keep it. For a fixed total $R$, the profile closest to white noise is the capped allocation (Lemma~\ref{lem:capped-profile}, Methods)
\begin{equation}
r^{\mathrm{cap}}_k(R)=\min\{\epsilon_k,\nu(R)\},
\qquad
\sum_{k=1}^L\min\{\epsilon_k,\nu(R)\}=R .
\label{eq:capped-allocation-main}
\end{equation}
It lowers the noisiest locations to a common level $\nu(R)$ and leaves the quieter ones untouched, so mitigation is spent only where the noise is concentrated. Capped PER then keeps the largest total noise that meets both requirements,
\begin{align}
R_{\max}&:=\min\Bigl\{\norm{\boldsymbol\epsilon}_1,\ \log\frac{1}{1-\gamma_\star}\Bigr\},
\label{eq:maximum-residual-budget}\\
R^\star&:=\max\Bigl\{R\in[0,R_{\max}]\ \Big|\ c_{\mathrm{scr}}\,e^{-R}\norm{\mathbf r^{\mathrm{cap}}(R)}_2\le\varepsilon_{\mathrm{tol}}\Bigr\},
\label{eq:optimal-residual-budget}
\end{align}
and implements the profile $\mathbf r^\star=\mathbf r^{\mathrm{cap}}(R^\star)$ with PER (Algorithm~\ref{alg:capped-per}, Methods). The result has effective level $\gamma_{\mathrm{out}}=1-e^{-R^\star}\le\gamma_\star$ and cost $\Gamma^\star=\exp[2(\norm{\boldsymbol\epsilon}_1-R^\star)]$. When the profile that reaches $\gamma_\star$ is too far from white noise, capped PER mitigates beyond the target, giving the validity of the white-noise description priority over cost.

Below the hardware noise level, the PER weights make the Born machine a quasi-probability model that is accessed through weighted expectation values rather than samples~\cite{vandenberg2023probabilistic,cai2023quantum}. The MMD estimates and gradients used in training are such expectations, which we obtain by executing many randomly sampled PER circuits and combining their outcomes with the known weights. At the hardware level no mitigation is applied, and the model can be sampled directly. Capped PER thus turns any target noise level into an implementable residual profile with a known cost, which leaves open the question of which level to target.

\subsection{Budget-aware noise-control regularization framework}

Capped PER realizes any target noise level, which leaves the second question of which level to target. The best level balances the contraction of the gap in Theorem~\ref{thm:effective-depolarizing-generalization} against the bias in equation~\eqref{eq:white-noise-bias} and the mitigation cost in equation~\eqref{eq:profile-cost-main}. This balance depends on the target distribution and on the model being trained, neither of which is known in advance, so the level cannot be fixed before training. The second component of \NAMEA{} therefore selects it during training, by comparing candidate levels on held-out data within a fixed budget of circuit executions. In this subsection, we formulate this selection problem, describe the controller that solves it, and bound the regret of its decisions.

\paragraph{Noise allocation problem.}
For a residual profile $\mathbf r$, let $R_{\mathbf r}(\theta):=\MMD_\kappa^2(p,q_\theta^{\mathbf r})$ denote the population loss under that profile. As discussed in the first subsection, this loss equals the training loss plus the generalization gap, so minimizing it accounts for both fit and generalization. Training performs $T$ parameter updates within a total budget of $B$ circuit executions. A policy $\pi$ chooses, at each step $t$, a residual profile $\mathbf r^\pi_t$ from the set $\mathcal R_{\mathrm{tol}}$ of profiles produced by capped PER and a number $N^\pi_t$ of shots per training circuit. At the comparison steps $t\in\mathcal I$, it also chooses a number $M^\pi_{tj}$ of probe shots for each candidate profile $j$. Its objective is
\begin{align}
\underset{\pi}{\operatorname{minimize}}\quad
&\mathbb E\!\left[
R_{\hat{\mathbf r}^{\pi}}(\hat\theta^{\pi})
\right]
\notag\\
\text{subject to}\quad
&\sum_{t=0}^{T-1} P_{\mathrm{circ}}N_t^\pi
+\sum_{t\in\mathcal I}\sum_{j\in\mathcal J_t}M_{tj}^\pi
\leq B,
\notag\\
&\mathbf r_t^\pi\in\mathcal R_{\mathrm{tol}},
\qquad t=0,\ldots,T-1.
\label{eq:budgeted-noise-control-problem}
\end{align}
where $P_{\mathrm{circ}}=1+2\dim(\theta)$ is the number of circuits per parameter-shift update. The constraint counts circuit executions, whereas the PER normalization determines how precise each execution is. A policy must therefore divide a single budget among the noise level, the precision of the training updates, and the comparisons between candidates.

\paragraph{Noise selection from held-out data.}
Because the population loss is unknown and depends on the whole training trajectory, we address equation~\eqref{eq:budgeted-noise-control-problem} with a practical policy that periodically compares candidate noise levels on held-out data (Algorithm~\ref{alg:noise-control} in Methods). Applying capped PER to a grid of target levels gives $J$ candidate profiles $\mathbf r^{(j)}$ with effective levels $\gamma_j=1-\exp(-\norm{\mathbf r^{(j)}}_1)$ and normalizations $\Gamma_j=\Gamma(\mathbf r^{(j)})$. A fraction $\alpha$ of the budget is reserved for comparisons and the rest for training, which is spread evenly so that each remaining training circuit receives $N_{\mathrm{future},t}$ shots. Every $K$ steps, the controller holds the parameters $\theta_t$ fixed, probes every candidate with $M_t$ fresh PER executions, and estimates its MMD to a held-out validation sample of size $v$ with a weighted U-statistic $\widetilde L^{U}_{\mathrm{val},t}(j)$ (Methods). It then selects
\begin{equation}
j_t\in\arg\min_{j\in\mathcal J}\Bigl\{
\underbrace{\widetilde L^{U}_{\mathrm{val},t}(j)}_{\text{validation fit}}
+\underbrace{b_{tj}}_{\text{measurement radius}}
+\underbrace{\lambda\,\frac{\Gamma_j^2}{N_{\mathrm{future},t}}}_{\text{budget lookahead}}
\Bigr\}
\label{eq:budget-aware-controller-score}
\end{equation}
and keeps the profile $\mathbf r^{(j_t)}$ until the next comparison. The validation fit is an empirical proxy for the population loss at the current parameters. The measurement radius $b_{tj}:=z\,\widehat{\mathrm{SE}}(\widetilde L^{U}_{\mathrm{val},t}(j))$, with $z\ge0$, keeps the controller from choosing a strongly mitigated candidate on the strength of a low but imprecise estimate. The budget lookahead penalizes candidates whose mitigation cost would leave too few shots for later training, and it relaxes as the remaining allowance grows. Training starts from the candidate with the largest $\gamma_j$, which is the cheapest, so noise is removed only when the validation evidence justifies its cost. Because the validation sample is reused across comparisons, final performance is assessed on an independent test set.

\paragraph{Cumulative selection regret.}
To assess these decisions, we compare the controller with the best candidate at every training step, with all candidates evaluated at the same current parameters,
\begin{equation}
\Psi_t(j):=R_{\mathbf r^{(j)}}(\theta_t)+\lambda\,\frac{\Gamma_j^2}{N_{\mathrm{future},t}},
\qquad
\mathrm{Reg}_T:=\sum_{t=0}^{T-1}\Bigl[\Psi_t(j_t)-\min_{j\in\mathcal J}\Psi_t(j)\Bigr].
\label{eq:regret-definition}
\end{equation}
The regret reflects both the accuracy of each comparison and the cost of keeping a profile while the parameters change. Its analysis rests on three facts. First, capped PER keeps every candidate within $\varepsilon_{\mathrm{tol}}$ of its white-noise model by equation~\eqref{eq:profile-dependent-scrambling}, so candidates differ mainly through their effective levels. Second, if each of the $d_\theta$ parameters enters once through a Pauli rotation, the loss changes with the parameters at a rate of at most $L_R:=2\sqrt{2d_\theta}$ (Lemma~\ref{lem:risk-lipschitz}, Methods). Third, the parameters themselves change slowly.

\begin{assumption}[Controlled updates]
\label{ass:controller-updates}
For a constant $h>0$ independent of $T$,
\begin{equation}
\|\theta_{t+1}-\theta_t\|_2\leq\frac{h}{t+1},
\qquad t=0,\ldots,T-1.
\label{eq:regret-update-assumption}
\end{equation}
\end{assumption}

Assumption~\ref{ass:controller-updates} holds for stochastic gradient descent with step size $\eta_0/(t+1)$ and gradient estimates bounded by $G$, with $h=\eta_0G$~\cite{bottou2018optimization}. It also holds approximately for Adam with the same decay, whose per-coordinate steps are bounded by a constant multiple of the learning rate~\cite{kingma2015adam}. The bound is stated in terms of a few quantities of the run. Comparisons occur at $\mathcal I=\{0,K,2K,\ldots\}\cap\{0,\ldots,T-1\}$, which gives $H=\lceil T/K\rceil$ rounds, and the comparison at step $s$ governs the next $w_s:=\min\{K,T-s\}$ steps. We write $\overline\Gamma:=\max_j\Gamma_j$ for the largest normalization, $\delta_\gamma:=\max_j\gamma_j-\min_j\gamma_j$ for the range of candidate levels, and $D_s:=\MMD_\kappa(q^0_{\theta_s},u)$ for how far the noiseless generator is from uniform at step $s$. Finally, $a_v(\beta):=v^{-1/2}+\sqrt{2\log(2/\beta)/v}$ bounds the error of the validation kernel mean embedding with probability at least $1-\beta/2$.

\begin{restatable}[Cumulative selection regret]{theorem}{theoremregret}
\label{thm:controller-regret}
Under Assumptions~\ref{ass:effective-depolarizing}
and~\ref{ass:controller-updates}, let $T\geq2$ and $1\leq K\leq T$.
Algorithm~\ref{alg:noise-control} satisfies, with probability at least
$1-\beta$,
\begin{align}
\operatorname{Reg}_T
\leq{}&
(15+6z)\,\overline\Gamma^2 T
\sqrt{\frac{JH\log(8JH/\beta)}{\alpha B}}
+2a_v(\beta)
\left(
\delta_\gamma\sum_{s\in\mathcal I}w_sD_s
+2T\varepsilon_{\mathrm{tol}}
\right)
\notag\\
&+2(K-1)
\left[
L_Rh(1+\log T)
+\frac{\lambda\overline\Gamma^2}{N_{\mathrm{future},0}}
\right].
\label{eq:regret-bound}
\end{align}
\end{restatable}

Theorem~\ref{thm:controller-regret} separates three sources of regret. The first term comes from comparing candidates with finite probe shots and from the measurement radius. It shrinks with the probe budget $\alpha B$ and grows with the number of candidates and comparison rounds, and its factor $\overline\Gamma^2$ shows that strongly mitigated candidates are expensive to compare, not only to train. The second term comes from the finite validation sample. It is scaled by how different the candidates actually are, so comparisons are robust when the generator is close to uniform or the candidate levels are similar. Because reusing the validation sample supplies no new target data, this term grows linearly in $T$ however many probe shots are spent. The last term is the price of holding a profile for $K$ steps while the loss and the training allowance change, and it vanishes for $K=1$. A larger probe fraction $\alpha$ sharpens comparisons but leaves fewer executions for training, which is the budget tradeoff of equation~\eqref{eq:budgeted-noise-control-problem}. Balancing the first and last terms through the comparison interval yields a sublinear rate.

\begin{restatable}[Rate under a linear execution budget]{corollary}{corollaryrate}
\label{cor:controller-regret-rate}
Suppose
$B=\Theta(T)$ and choose
$K=\lceil T^{2/3}\rceil$.
For fixed circuit size, candidate grid, $h$, $\alpha$, $\lambda$, $z$,
and $\varepsilon_{\mathrm{tol}}$, with probability at least $1-\beta$,
\begin{equation}
\operatorname{Reg}_T
=\widetilde{\mathcal O}\!\left(T^{2/3}+\frac{T}{\sqrt v}\right),
\label{eq:regret-sublinear-example}
\end{equation}
where $\widetilde{\mathcal O}$ suppresses logarithmic factors in
$T$ and $1/\beta$.
\end{restatable}

The interval $K=\lceil T^{2/3}\rceil$ balances the precision gained from less frequent comparisons against the cost of acting on stale ones. At fixed confidence, the average regret $\mathrm{Reg}_T/T$ then decays as $\mathcal O(T^{-1/3}\log T+v^{-1/2})$. The errors from finite probes and delayed comparisons therefore vanish as training proceeds, whereas a fixed validation set leaves a floor of order $v^{-1/2}$. Ultimately, it is the amount of held-out data, not the number of shots, that limits how well the noise level can be selected.

\begin{figure}[t]
\centering
\includegraphics[width=\textwidth]{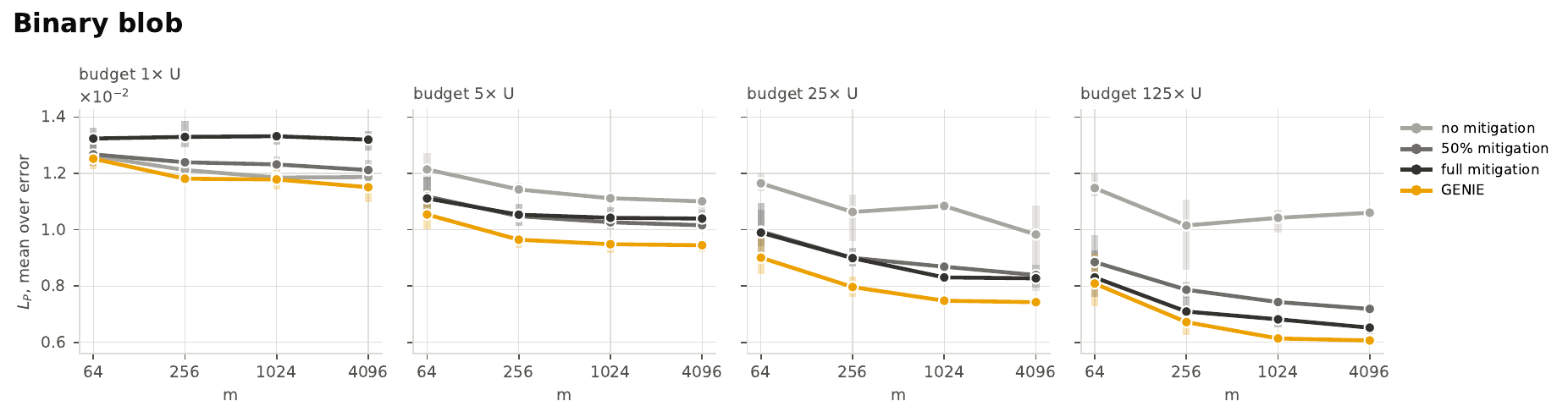}\\[2pt]
\includegraphics[width=\textwidth]{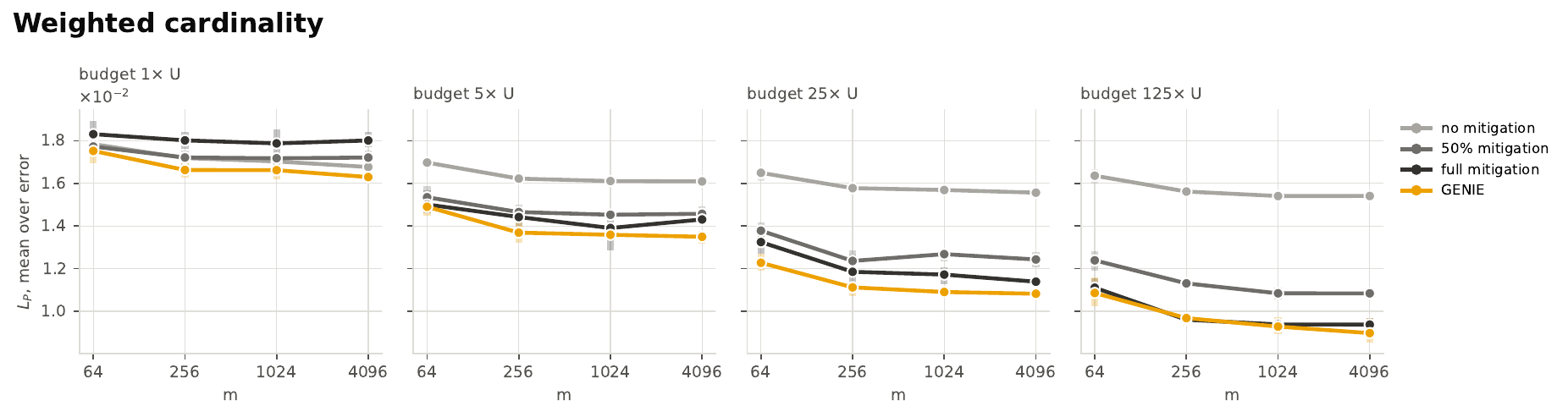}\\[2pt]
\includegraphics[width=\textwidth]{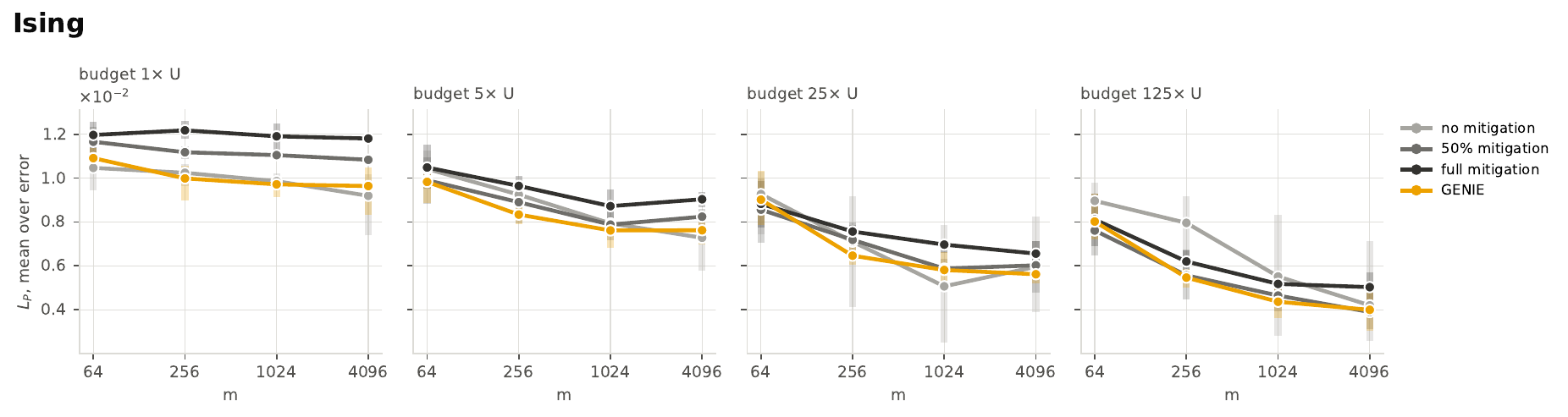}
\caption{\textbf{Final population loss across training-set sizes and execution budgets.} Rows show the blob (top), wfc (middle) and Ising (bottom) datasets, and panels within a row show the budgets $B\in\{1,5,25,125\}\times U$. Each curve shows the final population loss $L_P$ against the training-set size $m$ for training without mitigation, with half mitigation and with full mitigation, and for \NAMEA{}. Values are averaged over the three error rates and the three seeds, and bars show the range across seeds. The vertical axis is shared within each row.}
\label{fig:budget-panels}
\end{figure}

\subsection{Numerical Demonstrations}

\subsubsection{Experiments on Synthetic Datasets}
\paragraph{Loss across training-set sizes and execution budgets}.
 
Figure~\ref{fig:budget-panels} shows the final population loss $L_P$ of each method as a function of the training-set size $m$, with one panel per execution budget and the three error rates averaged. The vertical axis is shared across the four panels of each target, so reading a row from left to right shows how each method converts additional executions into a better model. Overall, \NAMEA{} gives the lowest loss of the four methods. On blob and wfc, it is the best method, or tied with the best within the seed range, in every panel. On Ising, it is among the best methods in every panel.
 
The clearest pattern is that the ranking of the fixed noise levels reverses as the budget grows. At the smallest budget, full mitigation is the worst on all three targets, because its PER estimates are $\Gamma^2$ times noisier and few executions cannot average this out. With more executions, the mitigated methods keep improving, while training without mitigation stalls on blob and wfc, because its noise-induced bias (equation~\eqref{eq:white-noise-bias}) does not shrink with more shots. Beyond the smallest budget, keeping all hardware noise is therefore the worst choice on these targets. No fixed level works well at every budget.

\NAMEA{} avoids this problem because it chooses the noise level from data. When executions are scarce, it keeps most of the hardware noise, as training without mitigation does. When executions are plentiful, it removes most of the noise, as full mitigation does. At the intermediate budgets, it does better than all three fixed levels, possibly because it can use levels between them or change the level during training. On Ising, the four methods are close and their seed ranges overlap, so the differences there are small.
 
Finally, the training-set size matters less than the budget. For most methods, the loss drops between $m=64$ and $m=256$ and changes little afterwards. On blob and wfc, the ranking of the methods also stays the same across $m$. In this range of $m$, the best noise level is therefore decided mainly by the budget and the error rate rather than by the amount of training data. Ising depends more strongly on $m$ at large budgets.
 

\paragraph{Regret across all conditions}
 
\begin{figure}[t]
\centering
\includegraphics[width=0.75\textwidth]{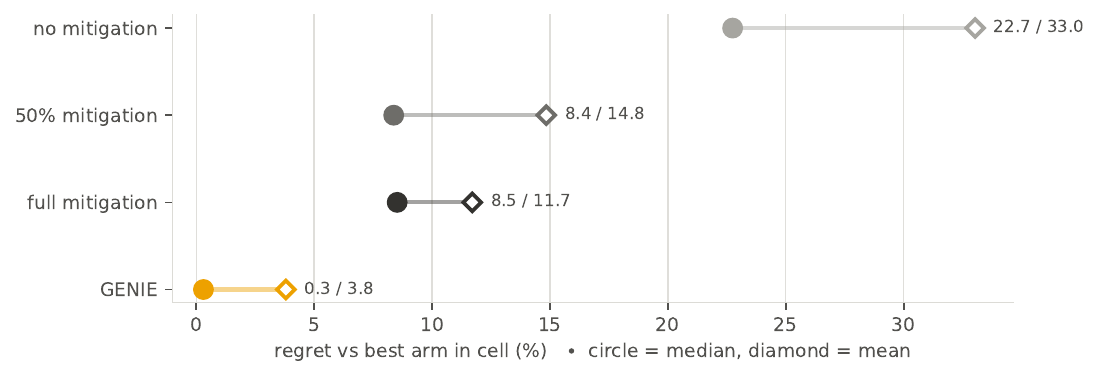}
\caption{\textbf{Regret of each method over all conditions.} Median (circles) and mean (diamonds) regret of each method relative to the best of the four methods, over the $144$ conditions of the grid, with losses averaged over three seeds before the regret is computed. Labels give the median and the mean.}
\label{fig:regret-summary}
\end{figure}

To compare the methods over the whole grid, we compute in each of the $144$ conditions (i.e., combinations of target, error rate, training-set size and budget). The regret is the percentage by which a method's loss exceeds that of the best of the four methods in that condition (Methods). It also differs from the cumulative selection regret $\mathrm{Reg}_T$ of Theorem~\ref{thm:controller-regret}, which sums, over the steps of a single training run, how much worse the level chosen by the controller is than the best candidate level. Figure~\ref{fig:regret-summary} summarizes the regrets.  \NAMEA{} has a median regret of $0.3\%$ and a mean regret of $3.8\%$, far below every fixed level. Full and half mitigation have median regrets of about $8.5\%$ and mean regrets of $11.7\%$ and $14.8\%$, and training without mitigation is the worst, with a median of $22.7\%$ and a mean of $33.0\%$. For every method the mean exceeds the median, because regret cannot be negative and a few large regrets pull the mean up.
 
\begin{figure}[t]
\centering
\begin{subfigure}[b]{0.555\textwidth}
\caption{}\label{fig:regret-profile-cdf}
\includegraphics[trim={0 0 360pt 0},clip,width=\textwidth]{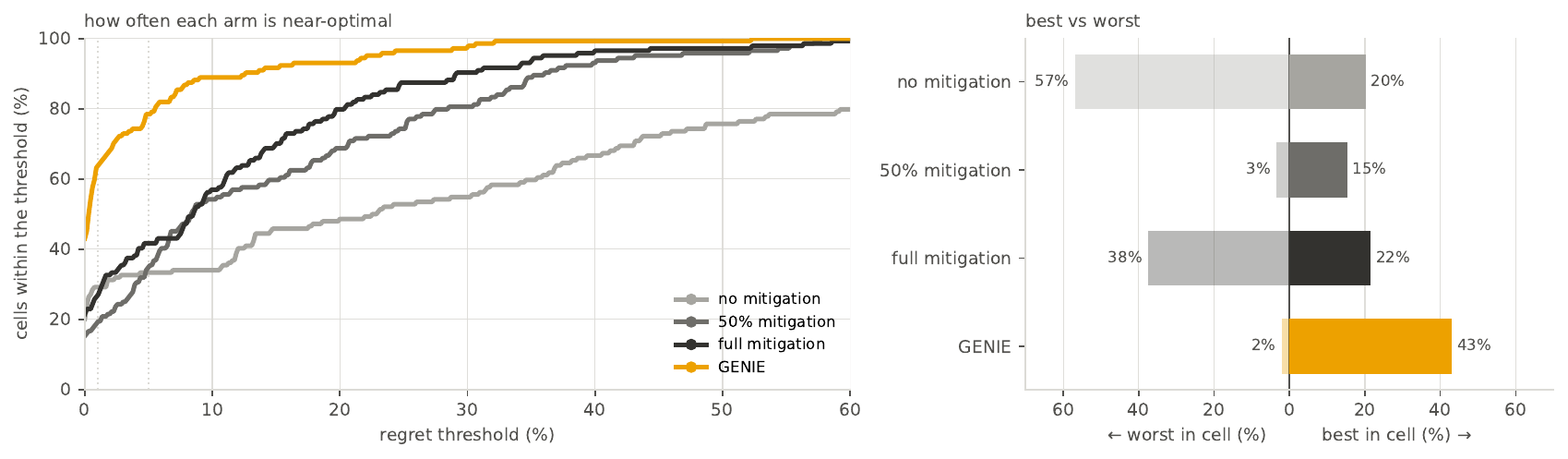}
\end{subfigure}\hfill
\begin{subfigure}[b]{0.435\textwidth}
\caption{}\label{fig:regret-profile-bestworst}
\includegraphics[trim={458pt 0 0 0},clip,width=\textwidth]{figs/regret_profile.pdf}
\end{subfigure}
\caption{\textbf{How often each method is near the best.} \textbf{a}, Percentage of the $144$ conditions in which each method is within a given regret threshold of the best method. A point $(x,y)$ means that the method is within $x\%$ of the best method in $y\%$ of the conditions. The dotted lines mark thresholds of $1\%$ and $5\%$. \textbf{b}, Percentage of conditions in which each method is the worst (left) and the best (right) of the four methods.}
\label{fig:regret-profile}
\end{figure}

 
Two summary numbers can hide how often a method fails, so Fig.~\ref{fig:regret-profile} shows the full picture. Panel~\textbf{a} plots, for each regret threshold $x$, the percentage of conditions in which a method is within $x\%$ of the best method. The curve of \NAMEA{} lies above those of all fixed levels at every threshold, so its advantage does not rest on a few favourable conditions. \NAMEA{} is within $1\%$ of the best method in $64\%$ of conditions and within $5\%$ in $78\%$, whereas no fixed level exceeds $29\%$ and $42\%$ at these thresholds.
 
Panel~\textbf{b} counts how often each method is the best and how often it is the worst in its condition. \NAMEA{} is the best method in $43\%$ of conditions and the worst in only $2\%$. The fixed levels are risky in different ways. Full mitigation has a moderate mean regret but is the worst method in $38\%$ of conditions, and training without mitigation is the worst in $57\%$. Half mitigation is rarely the worst ($3\%$) but also rarely the best ($15\%$).
 
Therefore, no fixed noise level is safe to use in practice. Each one performs badly in a sizeable fraction of conditions, and which conditions those are depends on the error rate and the budget at hand. In contrast, \NAMEA{} is not always the best method, but it is almost never far from the best.
 
\subsubsection{Experiments on a Real Census Dataset}
 
The synthetic experiments evaluate distributional fit through population MMD loss. To assess whether these improvements translate into more accurate population statistics and better predictions, we repeat the comparison on the Adult census dataset~\cite{becker1996adult}. Each $10$-bit string encodes a person's record through seven attributes, including whether their income exceeds \$50K (Methods). We evaluate the learned distributions against held-out real records using three practical criteria: the shares of people satisfying individual attributes, income rates within demographic subgroups, and the accuracy of income predictions for unseen individuals. These evaluations test whether lower MMD loss corresponds to better preservation of interpretable features of the data distribution. 

\begin{figure}[t]
\centering
\includegraphics[width=\textwidth]{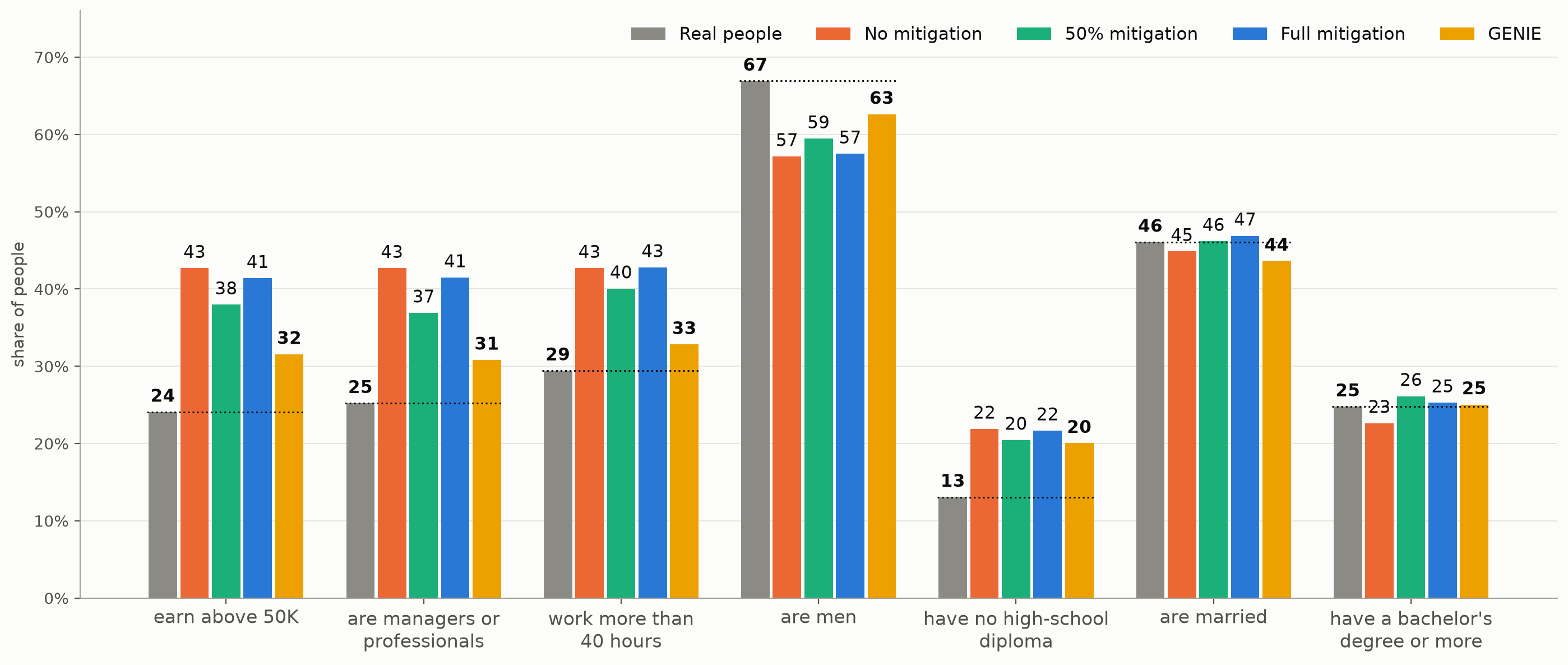}
\caption{\textbf{Generated population shares on Adult.}
Each group compares the percentage of records satisfying the indicated
event in held-out real data (gray) and samples generated by four methods (colored). Dotted lines mark the real-data percentages. Smaller absolute
gaps from these lines indicate more accurate marginal probabilities. Labels above bars show rounded percentages.
\NAMEA{} is closest or tied for closest on six of the seven events, with an average absolute discrepancy of approximately $4$ percentage
points, versus $8$--$10$ for the fixed mitigation levels.
Values are averaged over combinations of $B\in\{5,25,125\}\times U$ and training-set sizes $m \in \{64, 128, 256, 512\}$.}
\label{fig:adult-shares}
\end{figure}

\paragraph{Population shares.}
We first evaluate whether generated samples reproduce the marginal probabilities of seven demographic and socioeconomic events in the Adult dataset. For each event, such as earning above \$50K, we compare the fraction of generated records satisfying that event with the corresponding fraction in the held-out real data. Figure~\ref{fig:adult-shares} shows these fractions. In details, the gap between a method's bar and the dotted reference line is its absolute error for that event.

\NAMEA{} achieves the smallest error, or ties for the smallest, on six of the seven events. The exception is the share of married people, where \NAMEA{} is two points below the real share and the other methods are within one point. For example, without mitigation, $43\%$ of generated records have income above \$50K, compared with $24\%$ in the real data. \NAMEA{} reduces this share to $32\%$. Similarly, it reduces the generated share of managers or professionals from $43\%$ to $31\%$, closer to the real share of $25\%$. Averaged over the seven events, the absolute discrepancy is approximately $4$ percentage points for \NAMEA{}, compared with $8$--$10$ percentage points for the fixed mitigation levels. All methods closely match the shares of married people and people with a bachelor's degree or higher, so these two events provide little separation between methods.


\paragraph{Income rates within groups.}
Matching marginal population shares does not ensure that a model preserves relationships between attributes. Therefore, we evaluate conditional income rates. Specifically, the rate is the fraction of records earning above \$50K within a specified subgroup, such as married men or people aged $28$ or younger without a high-school diploma. We consider $125$ subgroups, each defined by a combination of values of two non-income attributes and containing at least $2\%$ of the real test records. For each subgroup $g$, we compare the generated income rate $\widehat q_g$ with the real test rate $\widehat p_g$. We summarize these discrepancies using the group-size-weighted mean absolute error,
$$
E_{\mathrm{rate}}
=
\frac{\sum_{g\in\mathcal G} n_g
      \left|\widehat q_g-\widehat p_g\right|}
     {\sum_{g\in\mathcal G} n_g},
$$
where $n_g$ is the number of real test records in subgroup $g$. The subgroups may overlap. The weighting gives greater importance to errors on larger groups.

\begin{figure}[t]
\centering
\includegraphics[width=\textwidth]{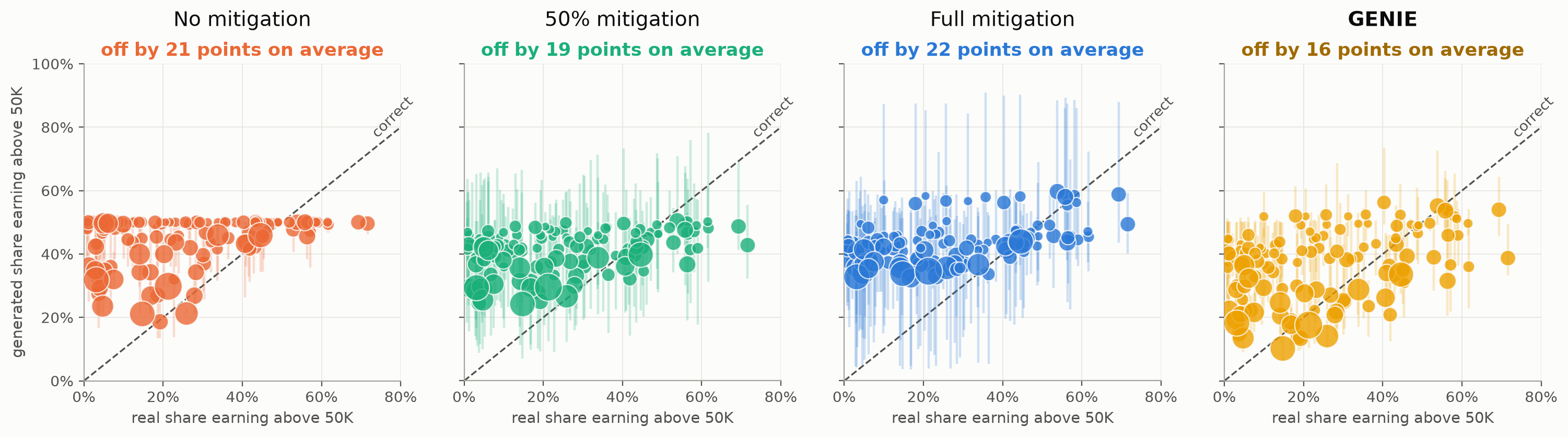}
\caption{\textbf{Preservation of subgroup income rates on Adult.}
Each marker represents one of $125$ subgroups defined by values of two
non-income attributes and covering at least $2\%$ of the real test records. Its horizontal coordinate is the fraction of real subgroup
members earning above \$50K, while its vertical coordinate is the corresponding fraction in generated data. The dashed diagonal denotes exact agreement such that points above it overestimate the subgroup's income rate, and points below it underestimate it. Marker size indicates real subgroup size. Vertical bars show how much one group's generated answer varies across runs. Panel annotations report the group-size-weighted mean absolute error in percentage points. \NAMEA{} reduces this error to $16$ points, compared with $19$--$22$ for the fixed mitigation levels.}
\label{fig:adult-rates}
\end{figure}

Figure~\ref{fig:adult-rates} plots the generated rate against the real
rate for each subgroup, with the diagonal indicating exact agreement.
The real rates range from approximately $0\%$ to $70\%$.
Without mitigation, many generated rates cluster near $50\%$,
including for groups whose real rates are much lower.
The fixed mitigation levels also overestimate income rates for
many low-income groups, while full mitigation exhibits substantial
variation across seeds.
\NAMEA{} produces lower rates for many of these groups and achieves the lowest weighted mean absolute error with $16$ percentage points, compared with $21$ without mitigation, $19$ with $50\%$ mitigation, and $22$ with full mitigation.

Nevertheless, \NAMEA{} does not fully recover the conditional income structure. Many low-income groups remain overestimated, and some high-income groups are underestimated. For context, the same metric is approximately $28$ percentage points for a uniform distribution over encoded records and less than
$1$ percentage point for the empirical training distribution.
Thus, \NAMEA{} improves conditional fidelity under the available measurement budget, but a gap remains relative to the relationships present in the training data.

\paragraph{Income prediction for real people.}
We next evaluate whether the learned distribution supports income prediction on unseen real records. For each test record, we condition on its nine non-income bits and use the model's conditional probability of the income bit to predict whether income exceeds \$50K, following the decision rule in Methods. This predictor is read directly from
the learned distribution.

\begin{figure}[t]
\centering
\includegraphics[width=\textwidth]{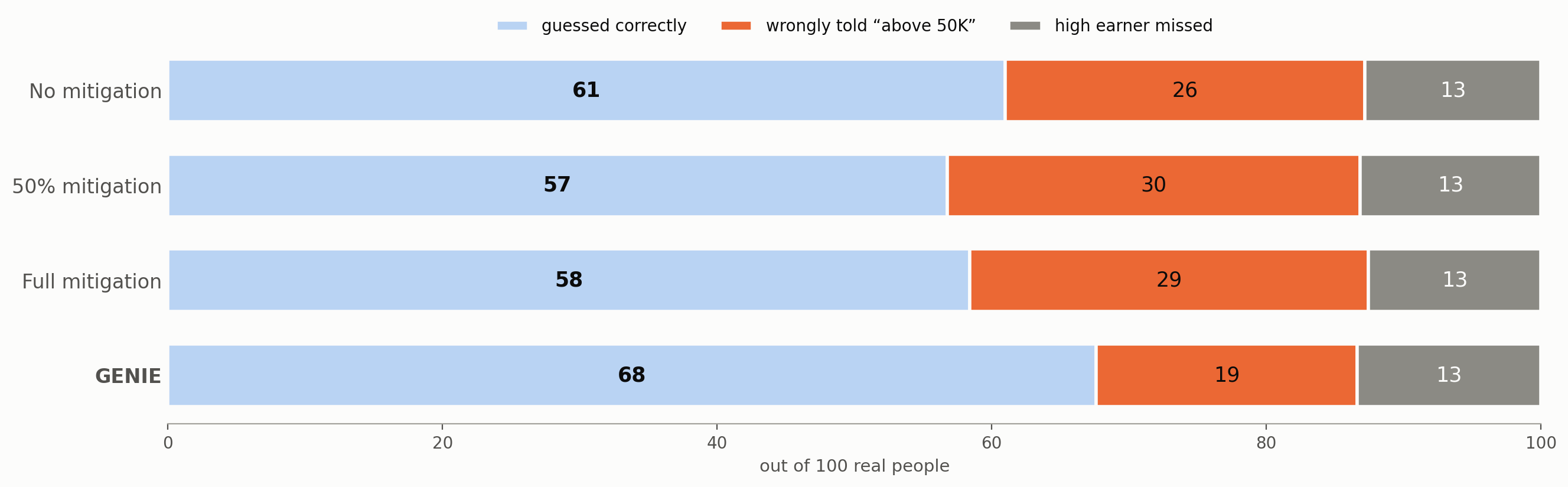}
\caption{\textbf{Income prediction on held-out records.}
Predictions are obtained by conditioning each learned distribution on a test record's nine non-income bits and applying the decision rule
in Methods. Each stacked bar represents $100$ real test records: blue denotes correct predictions, orange denotes false positives (income at most \$50K predicted as above \$50K), and gray denotes
false negatives (income above \$50K predicted as at most \$50K).}
\label{fig:adult-decisions}
\end{figure}

Figure~\ref{fig:adult-decisions} partitions the real test records into three outcomes including correct predictions, false positives (income at most \$50K predicted as above \$50K), and false negatives (income above \$50K predicted as at most \$50K). Each bar reports these outcomes per $100$ test records. \NAMEA{} correctly classifies approximately $68$ records, compared with $61$ without mitigation, $57$ with $50\%$ mitigation, and $58$ with full mitigation. This improvement comes primarily from fewer false positives, which are, $19$ per $100$ records for \NAMEA{}, versus $26$--$30$ for the fixed mitigation levels. All four methods produce approximately $13$ false negatives per $100$ records at the displayed precision.

\section{Discussion}

In this work, we have shown that the residual noise of a quantum circuit Born machine acts as a regularizer whose strength can be quantified, realized on hardware and chosen during training. The analysis rests on an identity that reduces the generalization gap of an MMD-trained Born machine to the overfitting of its witness function, so that for any noise process the gap is controlled by the complexity of the witness class (Theorem~\ref{thm:noise-agnostic-generalization}). For scrambled local Pauli noise, this complexity contracts by the retained-signal factor $1-\gamma$ (Theorem~\ref{thm:effective-depolarizing-generalization}). The best noise level can be intermediate and unknown before training. \NAMEA{} turns this trade-off into a procedure. Capped probabilistic error reduction realizes any target level at the lowest mitigation cost while keeping the remaining noise close to white noise, and a budget-aware controller selects the level from held-out data with an average regret per step that decays as $T^{-1/3}$ up to a floor of order $v^{-1/2}$ set by the held-out sample size (Theorem~\ref{thm:controller-regret} and Corollary~\ref{cor:controller-regret-rate}).

These results add a second consideration to how error mitigation is used. Mitigation is usually judged by how much bias it removes at a given sampling cost~\cite{cai2023quantum}. When a generative model is trained under a fixed budget of circuit executions, however, mitigation buys noisier shots rather than fewer of them, and the noise it removes no longer regularizes the model. How much noise to remove then becomes a model-selection problem that depends on the target, the training data and the budget. Our simulations show what is at stake. Across $144$ synthetic conditions, the fixed noise levels had median regrets of $8.4\%$ to $22.7\%$ relative to the best of the four methods, and each was the worst method in some conditions, whereas \NAMEA{} had a median regret of $0.3\%$. On the Adult census data at strong noise, it gave the most accurate population shares, income rates and income predictions of the four methods. This view parallels the use of injected noise as a regularizer in classical learning~\cite{bishop1995training,srivastava2014dropout} and the finite optimal noise levels found for supervised quantum models~\cite{heyraud2022noisy,bu2023effects,zhang2026theory}. For a Born machine, however, the noisy output distribution is itself the model, so the regularizer carries a bias that we quantify and that the controller weighs against the reduction of the gap.

Our study has three remaining limitations. First, the effective depolarizing approximation underlying Theorem~\ref{thm:effective-depolarizing-generalization} and the controller requires sufficient circuit scrambling to keep the approximation error small. Thus, its precision on estimating the amount of residual noise depends on the circuit architecture and depth. Second, the controller compares candidate noise levels at its current parameters rather than planning a schedule for the whole run, and the reliability of its choices depends on the held-out sample size. In the census experiment, held-out samples of $13$ to $102$ people were already sufficient for \NAMEA{} to select useful noise levels. Finally, all experiments use simulated circuits of $10$ and $12$ qubits.

These limitations suggest several natural extensions. Testing \NAMEA{} on larger circuits and hardware would help assess the effective depolarizing approximation and the benefits of noise control under realistic conditions. Randomized compiling and sparse noise learning provide tools for tailoring and characterizing the Pauli channels used by capped probabilistic error reduction~\cite{wallman2016noise,vandenberg2023probabilistic}, allowing the mismatch $\varepsilon_{\mathrm{Pauli}}$ to be assessed along training trajectories. On the theory side, Theorem~\ref{thm:noise-agnostic-generalization} covers arbitrary noise processes, while the explicit contraction in Theorem~\ref{thm:effective-depolarizing-generalization} relies on scrambled Pauli noise. Deriving contraction bounds for other channels would clarify which correlations they suppress and when retaining noise is beneficial. More broadly, our results suggest that residual noise, when appropriately characterized and controlled, can serve as a resource for generalization on near-term hardware.

\section{Methods}

This section collects the technical ingredients behind the Results. We first give the lemmas and proof sketches behind the generalization bounds, then the construction of capped probabilistic error reduction, then the implementation and analysis of the budget-aware controller, and finally the setup of our numerical experiments. Complete proofs are given in Supplementary Note~\tbd{1}.

\subsection{Generalization bounds}

\paragraph{Reduction to the learned witness.}
Both generalization theorems start from an exact identity, which shows that only the coupling term of the MMD can make the empirical risk optimistic.

\begin{restatable}[Reduction to the learned witness]
{lemma}{lemmaboundtowitness}
\label{lem:expected-gap-identity}
Let $m,N\ge 2$, and let $\A_{\mathcal N}$ be any possibly randomized learning
algorithm. If the evaluation samples $\Y$ are conditionally independent of
$\Ss$ and the training randomness given $\hat\theta_{\mathcal N}$, then
\begin{equation}
\Delta_{\mathcal N}(m)
=
2\,
\mathbb E_{\Ss,\A_{\mathcal N}}
\left[
(P_m-P)h_{\hat\theta_{\mathcal N}}^{\mathcal N}
\right].
\label{eq:gap-identity}
\end{equation}
\end{restatable}

\textit{Proof sketch.} The target--target U-statistic is unbiased and independent of $\theta$. The model--model U-statistic uses samples drawn after training and is therefore unbiased conditionally on $\hat\theta_{\mathcal N}$. In the coupling term, averaging over $\Y$ given $\Ss$ and $\hat\theta_{\mathcal N}$ replaces $\frac1N\sum_j\kappa(x_i,y_j)$ by $h^{\mathcal N}_{\hat\theta_{\mathcal N}}(x_i)$, so the sample coupling term becomes $P_mh^{\mathcal N}_{\hat\theta_{\mathcal N}}$. Its population counterpart is $Ph^{\mathcal N}_{\hat\theta_{\mathcal N}}$, which equals the coupling term of the risk by the definition of the witness, equation~\eqref{eq:prelim-witness}. Only the deviation in equation~\eqref{eq:gap-identity} remains, and it does not involve $N$.

\textit{Proof sketch of Theorem~\ref{thm:noise-agnostic-generalization}.} The learned witness belongs to $\W^{\mathcal N}$, so Lemma~\ref{lem:expected-gap-identity} gives $|\Delta_{\mathcal N}(m)|\le2\,\mathbb E_{\Ss}\sup_{h\in\W^{\mathcal N}}|(P_m-P)h|$. The standard symmetrization argument~\cite{bartlett2002rademacher} bounds this uniform deviation by the Rademacher complexity,
\begin{equation}
\mathbb E_{\Ss}\Bigl[\sup_{h\in\W^{\mathcal N}}\bigl|(P_m-P)h\bigr|\Bigr]\le2\,\Rad\bigl(\W^{\mathcal N}_\pm\bigr),
\label{eq:uniform-deviation}
\end{equation}
which gives $|\Delta_{\mathcal N}(m)|\le4\Rad(\W^{\mathcal N}_\pm)$.

\paragraph{Complexity of the noiseless witness class.}
Theorem~\ref{thm:effective-depolarizing-generalization} reduces the noisy complexity to that of the centered noiseless witnesses $\bar h_\theta:=h^0_\theta-\widehat\phi(\varnothing)$ and their class $\bar\W^0_\pm:=\{\pm\bar h_\theta:\theta\in\Theta\}$, which the parity basis makes explicit.

\begin{restatable}[Complexity of the noiseless witness class]
{lemma}{lemmanoiseless}
\label{lem:rademacher-anticoncentration}
Under Assumption~\ref{ass:translation-invariant-kernel}, and for a given $C\geq1$ such that $\sup_{\theta\in\Theta}
2^n\sum_{x\in\X}q_\theta^0(x)^2
\leq
C$, we have:
\begin{equation}
\Rad\bigl(\bar{\W}_\pm^0\bigr)
\leq
\frac{\Lambda_C}{\sqrt m},
\label{eq:population-rademacher-bound}
\end{equation}
where $\Lambda_C$ is the constant defined in Theorem~\ref{thm:effective-depolarizing-generalization}.
\end{restatable}

\textit{Proof sketch.} By equation~\eqref{eq:witness-fourier}, the Rademacher average of a centered witness is a sum over nonconstant parities,
\begin{equation*}
\frac1m\sum_{i=1}^m\sigma_i\bar h_\theta(x_i)=\sum_{A\neq\varnothing}\widehat\phi(A)\,\widehat{q^0_\theta}(A)\,g_A,
\qquad
g_A:=\frac1m\sum_{i=1}^m\sigma_i\chi_A(x_i).
\end{equation*}
The Cauchy--Schwarz inequality over $A$ separates the generator from the kernel and the sample. Parseval's identity gives $\sum_{A\neq\varnothing}\widehat{q^0_\theta}(A)^2=2^n\sum_xq^0_\theta(x)^2-1\le C-1$, and Jensen's inequality with $\mathbb E_\sigma g_A^2=1/m$ bounds the remaining factor by $(\sum_{A\neq\varnothing}\widehat\phi(A)^2/m)^{1/2}$. Weighting the Cauchy--Schwarz inequality by $\widehat\phi(A)$ instead, and using $|\widehat{q^0_\theta}(A)|\le1$ and $\sum_A\widehat\phi(A)=\phi(\mathbf 0)=1$, gives the second branch $1-\widehat\phi(\varnothing)$ of $\Lambda_C$.

\textit{Proof sketch of Theorem~\ref{thm:effective-depolarizing-generalization}.} The proof combines Lemma~\ref{lem:expected-gap-identity} with Lemma~\ref{lem:rademacher-anticoncentration}. Because $\widehat u(A)=0$ for every $A\neq\varnothing$, the white-noise witness is $h^\gamma_\theta=\widehat\phi(\varnothing)+(1-\gamma)\bar h_\theta$. Writing $e_\theta:=h^{\mathcal N}_\theta-h^\gamma_\theta=\mu_{q^{\mathcal N}_\theta}-\mu_{q^\gamma_\theta}$ for the remainder, every physical witness decomposes as
\begin{align}
h_\theta^{\mathcal N}&=\widehat\phi(\varnothing)+(1-\gamma)\,\bar h_\theta+e_\theta ,
\label{eq:physical-effective-witness-decomposition}\\
\norm{e_\theta}_{\Hil}&=\MMD_\kappa\bigl(q_\theta^{\mathcal N},q_\theta^\gamma\bigr)\le\varepsilon_{\mathrm{Pauli}},
\label{eq:witness-effective-error-bound}
\end{align}
where the inequality is Assumption~\ref{ass:effective-depolarizing}. Inserting this decomposition into equation~\eqref{eq:gap-identity}, the constant vanishes by equation~\eqref{eq:prelim-centering}, and the contracted component contributes at most $4(1-\gamma)\Rad(\bar{\W}^0_\pm)$. The remainder satisfies $(P_m-P)e_\theta=\inner{e_\theta}{\frac1m\sum_i\kappa(x_i,\cdot)-\mu_p}_{\Hil}\le\varepsilon_{\mathrm{Pauli}}\norm{\frac1m\sum_i\kappa(x_i,\cdot)-\mu_p}_{\Hil}$, and the expected error of the empirical kernel mean is at most $m^{-1/2}$, which gives the term $2\varepsilon_{\mathrm{Pauli}}/\sqrt m$.

\textit{Proof sketch of Corollary~\ref{cor:effective-depolarizing-high-probability}.} McDiarmid's inequality controls how far the supremum over the contracted witness class and the error of the empirical kernel mean deviate from their expectations, and both deviations inherit the factor $1-\gamma$. Hoeffding's inequality for U-statistics controls the target--target estimate and the terms that involve the evaluation samples $\Y$. A union bound over these events gives the stated probability.

\subsection{Capped probabilistic error reduction}

This subsection gives the quasi-probability construction behind the cost in equation~\eqref{eq:profile-cost-main} and the procedure that computes the capped profile.

\paragraph{Quasi-probability implementation.}
At location $k$, the residual channel with rate $r_k\le\epsilon_k$ keeps the Pauli distribution of the characterized channel~\eqref{eq:characterized-pauli-channel},
\begin{equation}
\mathcal E_{k,r_k}(\rho)=(1-r_k)\rho+r_k\sum_{P\in\{X,Y,Z\}}\pi_k(P)\,P\rho P .
\label{eq:residual-pauli-channel}
\end{equation}
PER implements the correction $\mathcal M_{k,r_k}:=\mathcal E_{k,r_k}\circ\mathcal E_k^{-1}$. This map is generally not a physical channel, but it can be written as a signed combination of Pauli conjugations, $\mathcal M_{k,r_k}=\sum_P\eta_{k,r_k}(P)\,\mathcal P$ with $\mathcal P(\rho)=P\rho P$. PER samples $\mathcal P$ with probability proportional to $|\eta_{k,r_k}(P)|$ and assigns the corresponding outcome the sign of $\eta_{k,r_k}(P)$. Because the corrections are sampled independently across locations, the circuit-level normalization is
\begin{equation}
\Gamma(\mathbf r):=\prod_{k=1}^L\Gamma_k(r_k),
\qquad
\Gamma_k(r_k):=\sum_P\bigl|\eta_{k,r_k}(P)\bigr|,
\label{eq:per-normalization}
\end{equation}
and every outcome carries a weight of magnitude $\Gamma(\mathbf r)$. The following lemma gives the cost used in the Results.

\begin{restatable}[Cost of a mitigation profile]{lemma}{lemmaprofilecost}
\label{lem:profile-cost}
Suppose that $\epsilon_k<1/2$ and
$0\leq r_k\leq\epsilon_k$ for every $k$. The circuit-level
quasiprobability normalization required to reduce
$\boldsymbol\epsilon$ to $\mathbf r$ satisfies
\begin{equation}
\log\Gamma(\mathbf r)
=
2\norm{\boldsymbol\epsilon-\mathbf r}_1
+
\mathcal O\left(
\sum_{k=1}^L
\epsilon_k(\epsilon_k-r_k)
\right).
\label{eq:profile-cost}
\end{equation}
Consequently, in the weak-noise regime,
\begin{equation}
\Gamma(\mathbf r)
\simeq
\exp\left(
2\norm{\boldsymbol\epsilon-\mathbf r}_1
\right).
\label{eq:leading-profile-cost}
\end{equation}
\end{restatable}

\textit{Proof sketch.} Each $\mathcal M_{k,r_k}$ is diagonal in the Pauli transfer basis. To first order in the rates, $\mathcal M_{k,r_k}\approx\mathrm{id}-(\epsilon_k-r_k)\sum_P\pi_k(P)(\mathcal P-\mathrm{id})$, so $\Gamma_k(r_k)=1+2(\epsilon_k-r_k)+\mathcal O(\epsilon_k(\epsilon_k-r_k))$. Taking logarithms and summing over the independent locations gives equation~\eqref{eq:profile-cost}. The model--model term of the MMD multiplies two such weights, so its variance has a leading term of order $\Gamma(\mathbf r)^2/N$ from fluctuations of one weight at a time and a term of order $\Gamma(\mathbf r)^4/N^2$ from both weights together

\paragraph{Capped allocation and the retained budget.}
The capped profile in equation~\eqref{eq:capped-allocation-main} solves the allocation step of problem~\eqref{eq:capped-per-design} at a fixed total residual noise.

\begin{restatable}[Capped allocation at a fixed budget]{lemma}
{lemmacappedprofile}
\label{lem:capped-profile}
Fix $R\in[0,\norm{\boldsymbol\epsilon}_1]$. The unique solution of
\begin{align}
\min_{\mathbf r}\quad&
\norm{\mathbf r}_2
\notag\\
\text{subject to}\quad&
\norm{\mathbf r}_1=R,
\qquad
0\leq r_k\leq\epsilon_k
\label{eq:fixed-budget-allocation}
\end{align}
is
\begin{equation}
r_k^{\mathrm{cap}}(R)
=
\min\{\epsilon_k,\nu(R)\},
\label{eq:capped-allocation}
\end{equation}
where $\nu(R)\geq0$ is chosen such that
\begin{equation}
\sum_{k=1}^L
\min\{\epsilon_k,\nu(R)\}
=
R.
\label{eq:capped-budget}
\end{equation}
\end{restatable}

\textit{Proof sketch.} The problem minimizes a strictly convex function over a polytope. Its Karush--Kuhn--Tucker conditions give $r_k=\min\{\epsilon_k,\max\{0,\nu\}\}$ for a multiplier $\nu$. Because $\sum_k\min\{\epsilon_k,\nu\}$ is continuous and nondecreasing in $\nu$, a value $\nu(R)$ satisfying equation~\eqref{eq:capped-budget} exists, and strict convexity gives uniqueness.

It remains to choose the retained budget $R^\star$ in equation~\eqref{eq:optimal-residual-budget}. Let $g(R):=e^{-R}\norm{\mathbf r^{\mathrm{cap}}(R)}_2$ denote the mismatch bound of the capped profile. Increasing $R$ raises only the capped entries, so $\frac{d}{dR}\norm{\mathbf r^{\mathrm{cap}}(R)}_2^2=2\nu(R)$ and $g'(R)$ has the sign of $\nu(R)-\norm{\mathbf r^{\mathrm{cap}}(R)}_2^2$. Since every entry is at most $\nu(R)$, this quantity is at least $\nu(R)(1-R)$, which is positive in the weak-noise regime $R<1$. The mismatch bound therefore increases with $R$, the feasible budgets form an interval, and $R^\star$ is found by bisection. Algorithm~\ref{alg:capped-per} summarizes the procedure.

\begin{algorithm}[t]
\caption{Capped probabilistic error reduction}
\label{alg:capped-per}
\begin{algorithmic}[1]
\Require characterized rates $\boldsymbol\epsilon$, target
$\gamma_\star$, tolerance $\varepsilon_{\mathrm{tol}}$, and scrambling
constant $c_{\mathrm{scr}}$
\Ensure residual profile $\mathbf r^\star$, realized level
$\gamma_{\mathrm{out}}$, and normalization $\Gamma^\star$
\State Compute $R_{\max}$ using
Equation~\eqref{eq:maximum-residual-budget}
\State Sort $\boldsymbol\epsilon$ to evaluate the capped profiles efficiently
\State Find $R^\star$ using Equation~\eqref{eq:optimal-residual-budget}
\State Find $\nu(R^\star)$ satisfying
Equation~\eqref{eq:capped-budget}
\State Set
$r_k^\star\gets\min\{\epsilon_k,\nu(R^\star)\}$ for every $k$
\For{$k=1,\ldots,L$}
    \State Configure
    $\mathcal M_{k,r_k^\star}
    \gets\mathcal E_{k,r_k^\star}\circ\mathcal E_k^{-1}$
\EndFor
\State $\gamma_{\mathrm{out}}\gets1-e^{-R^\star}$
\State
$\Gamma^\star\gets
\exp\left[
2(\norm{\boldsymbol\epsilon}_1-R^\star)
\right]$
\State \Return
$\mathbf r^\star,\gamma_{\mathrm{out}},\Gamma^\star$
\end{algorithmic}
\end{algorithm}

\subsection{Budget-aware noise control}

This subsection specifies the shot allocation and the validation estimator used by the \NAMEA{} controller, and then sketches the analysis behind Theorem~\ref{thm:controller-regret} and Corollary~\ref{cor:controller-regret-rate}.

\paragraph{Shot allocation and validation estimates.}
Let $B_{\mathrm{tr},t}$ and $B_{\mathrm{pr},t}$ denote the unspent training and probe budgets at step $t$, initialized to $(1-\alpha)B$ and $\alpha B$. Each training step receives an equal share of the remaining training budget, and each comparison an equal share of the remaining probe budget,
\begin{equation}
N_{\mathrm{future},t}:=\frac{B_{\mathrm{tr},t}}{P_{\mathrm{circ}}(T-t)},
\qquad
N_t:=\bigl\lfloor N_{\mathrm{future},t}\bigr\rfloor,
\qquad
M_t:=\Bigl\lfloor\frac{B_{\mathrm{pr},t}}{J\,u_t}\Bigr\rfloor,
\label{eq:controller-future-shots}
\end{equation}
where $u_t$ is the number of comparisons remaining, including the current one. Subtracting the executions spent at each step keeps the policy within the budget of equation~\eqref{eq:budgeted-noise-control-problem}. At a comparison, the PER probes of candidate $j$ return independent outcome--weight pairs $(Y_{tj,a},W_{tj,a})$ with $|W_{tj,a}|\le\Gamma_j$ and $\mathbb E[W_{tj,a}\kappa(Y_{tj,a},\cdot)]=\mu_{q^{\mathbf r^{(j)}}_{\theta_t}}$. On the validation sample $\Ss_{\mathrm{val}}=\{x_1,\ldots,x_v\}$, the controller uses the weighted U-statistic
\begin{equation}
\begin{split}
\widetilde L^{U}_{\mathrm{val},t}(j)
={}&\frac{1}{v(v-1)}\sum_{i\neq i'}\kappa(x_i,x_{i'})
+\frac{1}{M_t(M_t-1)}\sum_{a\neq b}W_{tj,a}W_{tj,b}\,\kappa(Y_{tj,a},Y_{tj,b})\\
&-\frac{2}{vM_t}\sum_{i=1}^v\sum_{a=1}^{M_t}W_{tj,a}\,\kappa(x_i,Y_{tj,a}),
\end{split}
\label{eq:controller-validation-ustat}
\end{equation}
which, averaged over the fresh probes, recovers the validation U-statistic with exact generator expectations. Excluding the diagonal pairs removes the self-interaction bias, and the estimates are compared without clipping because a U-statistic can be negative. The standard error in the measurement radius is estimated conditionally on the validation data and accounts for the dependence among the pairwise terms. Algorithm~\ref{alg:noise-control} combines these rules with the selection rule~\eqref{eq:budget-aware-controller-score}.

\begin{algorithm}[t]
\caption{Budget-aware noise-controlled regularization}
\label{alg:noise-control}
\begin{algorithmic}[1]
\Require Target data $\Ss$, fixed kernel $\kappa$, characterized hardware,
budget $B$, steps $T$, update interval $1\leq K<T$, target noise grid,
mismatch tolerance $\varepsilon_{\mathrm{tol}}$, and $\alpha,z,\lambda$
\Ensure Trained parameters $\hat\theta$ and residual profile $\hat{\mathbf r}$
\State Split $\Ss$ into $\Ss_{\mathrm{tr}}$ and $\Ss_{\mathrm{val}}$
\State Construct $J$ distinct admissible profiles
$\mathbf r^{(1)},\ldots,\mathbf r^{(J)}$ using
Algorithm~\ref{alg:capped-per}; record $\gamma_j,\Gamma_j$
\State $\mathcal J\gets\{1,\ldots,J\}$;
$\mathcal I\gets\{0,K,2K,\ldots\}\cap\{0,\ldots,T-1\}$
\State Initialize $\theta_0$; set $\mathbf r$ to the candidate with
largest $\gamma_j$
\State $P_{\mathrm{circ}}\gets1+2\dim(\theta)$;
$B_{\mathrm{tr},0}\gets(1-\alpha)B$;
$B_{\mathrm{pr},0}\gets\alpha B$
\For{$t=0,\ldots,T-1$}
    \State Compute $N_{\mathrm{future},t}$ and $N_t$ using
    Equation~\eqref{eq:controller-future-shots}; set $M_t\gets0$
    \If{$t\in\mathcal I$}
        \State $u_t\gets|\{s\in\mathcal I:s\geq t\}|$;
        compute $M_t$ using Equation~\eqref{eq:controller-future-shots}
        \For{$j\in\mathcal J$}
            \State At fixed $\theta_t$, collect $M_{tj}=M_t$ fresh PER
            probe shots under profile $\mathbf r^{(j)}$
            \State Compute $\widetilde L_{\mathrm{val},t}^{U}(j)$ and $b_{tj}$
            using Equation~\eqref{eq:controller-validation-ustat}
        \EndFor
        \State Select $j_t$ using
        Equation~\eqref{eq:budget-aware-controller-score};
        set $\mathbf r\gets\mathbf r^{(j_t)}$
    \EndIf
    \State $\mathbf r_t\gets\mathbf r$
    \State Estimate the training MMD gradient using
    $\Ss_{\mathrm{tr}}$, profile $\mathbf r_t$, and $N_t$ shots per circuit
    \State Update $\theta_t$ to $\theta_{t+1}$
    \State $B_{\mathrm{tr},t+1}\gets
    B_{\mathrm{tr},t}-P_{\mathrm{circ}}N_t$
    \State $B_{\mathrm{pr},t+1}\gets B_{\mathrm{pr},t}-JM_t$
\EndFor
\State \Return $\hat\theta\gets\theta_T$,
$\hat{\mathbf r}\gets\mathbf r$, and
$\hat\gamma=1-\exp(-\|\hat{\mathbf r}\|_1)$
\end{algorithmic}
\end{algorithm}

\paragraph{Regret analysis.}
The regret bound needs control over how fast candidate scores change between comparisons. On the circuit side, this is provided by the following sensitivity bound.

\begin{restatable}[Risk sensitivity]{lemma}{lemmarisklipschitz}
\label{lem:risk-lipschitz}
Suppose each of the $d_\theta=\dim(\theta)$ parameters occurs once,
through a Pauli rotation $e^{-i\theta_kP_k/2}$, with fixed,
parameter-independent CPTP noise channels and fixed measurement.
Then, uniformly over admissible residual profiles,
\begin{equation}
\left|R_{\mathbf r}(\theta)-R_{\mathbf r}(\theta')\right|
\leq L_R\|\theta-\theta'\|_2,
\qquad L_R:=2\sqrt{2d_\theta}.
\label{eq:regret-risk-lipschitz}
\end{equation}
\end{restatable}

\textit{Proof sketch.} The parameter-shift rule writes each partial derivative of the kernel mean embedding $\mu_{q^{\mathbf r}_\theta}$ as half the difference of the embeddings of two valid output distributions. Because every MMD is at most $\sqrt2$, each partial derivative of $R_{\mathbf r}$ is bounded by a constant, and summing over the $d_\theta$ coordinates gives the bound.

\textit{Proof sketch of Theorem~\ref{thm:controller-regret}.} Group the training steps by the comparison that governs them, and split the regret of each step into the selection error at that comparison and the change in scores since then. The selection error has three parts. The first is the probe error, which is controlled by the concentration of the weighted U-statistics with weights bounded by $\overline\Gamma$ and a union bound over the $JH$ estimates, and it gives the first term. The second is the measurement radius, which is at most $6z\overline\Gamma^2/\sqrt{M_s}$ because the standard error is bounded by the range of its per-shot terms, and it adds $6z$ to the constant of the first term. The third is the validation error, which is controlled by $a_v(\beta)$ and by the distance $\delta_\gamma D_s+2\varepsilon_{\mathrm{tol}}$ between candidate mean embeddings implied by equation~\eqref{eq:profile-dependent-scrambling}, and it gives the second term. The change in scores within each block is bounded with Lemma~\ref{lem:risk-lipschitz} and Assumption~\ref{ass:controller-updates} by summing the step sizes $h/(t+1)$, and together with the change of the budget penalty it gives the last term.

\textit{Proof sketch of Corollary~\ref{cor:controller-regret-rate}.} With $B=\Theta(T)$ and $K=\lceil T^{2/3}\rceil$, there are $H=\mathcal O(T^{1/3})$ comparisons, so the first term of equation~\eqref{eq:regret-bound} is $\widetilde{\mathcal O}(T^{2/3})$. The second term is $\widetilde{\mathcal O}(T/\sqrt v)$ because $D_s\le\sqrt2$. The last term is $\mathcal O(T^{2/3}\log T)$ because $N_{\mathrm{future},0}=(1-\alpha)B/(P_{\mathrm{circ}}T)=\Theta(1)$.

\subsection{Experimental Setup}
 
We evaluate \NAMEA{} in simulation on three synthetic targets at $n=12$ qubits and one real dataset (i.e., UCI Adult Census~\cite{becker1996adult}), varying the training-set size, the hardware error rate and the execution budget. We first describe the setup for accessment on synthetic datasets first and spend the last paragraph on the real census experiment.
 
\paragraph{Synthetic Datasets.}
The first is a binary blob distribution generated by eight prototype strings whose bits are flipped independently with probability $0.2$. The second is a weighted fixed-cardinality distribution on the $\binom{12}{6}=924$ strings of Hamming weight six, with weight parameter $\omega=6$. The third is the Boltzmann distribution of a ferromagnetic Ising ring at temperature $2.4$. Their effective numbers of outcomes, $e^{H(p)}$ with $H$ the Shannon entropy, are $1626$, $560$ and $1324$. In each run we draw $m\in\{64,256,1024,4096\}$ samples from the target and hold out $20\%$ of them as the validation sample of the controller.
 
\paragraph{Model, loss and noise.}
The Born machine is a hardware-efficient circuit with $10$ layers of single-qubit Pauli rotations and $\text{CNOT}$, giving $360$ parameters that each enter once through a Pauli rotation. Gradients are computed with the parameter-shift rule, so each update evaluates $P_{\mathrm{circ}}=721$ circuits. The loss is the MMD U-statistic with the Gaussian mixture kernel on the Hamming distance, with bandwidths $(0.5,1,2)\times\sqrt{n/5}$. Hardware noise consists of stochastic Pauli channels at the $L=240$ noisy locations of the circuit, with per-location rates of mean $\bar\epsilon\in\{0.002,0.004,0.008\}$ shared by all runs. Simulations are implemented in JAX and run on CPUs.
 
\paragraph{Training, budgets and compared methods.}
Each run performs $T=500$ updates with the Adam optimizer at the starting learning rate of $0.1$ which decays over time. Budgets are counted in circuit executions without any rescaling by the PER cost. The unit budget $U=T\,P_{\mathrm{circ}}\times2000\approx7.2\times10^8$ corresponds to $2000$ shots per circuit at every step, and we consider $B\in\{1,5,25,125\}\times U$, with each step receiving an even share of the remaining budget as in equation~\eqref{eq:controller-future-shots}. We compare four methods at the same budget. Three keep a fixed noise level throughout training, namely no mitigation ($\gamma=\bar\gamma$), half mitigation ($\gamma=\bar\gamma/2$) and full mitigation ($\gamma=0$). The fourth is \NAMEA{}, which compares the candidate levels $\gamma/\bar\gamma\in\{0,0.25,0.5,0.75,1\}$ every $K=25$ steps, starting at the first step, for $20$ comparisons per run. It uses $z=1$, $\lambda=1$ and a probe fraction $\alpha=0.02$, and its probe executions are charged to its own budget. These controller settings were tuned once at $n=10$ and transferred unchanged.
 
\paragraph{Evaluation and statistics.}
The grid comprises $3$ targets, $3$ error rates, $4$ sample sizes and $4$ budgets, giving $144$ conditions, each run with three seeds for a total of $1728$ runs. The figure of merit is the exact population loss $L_P=\MMD^2_\kappa(p,q)$ of the output distribution $q$ of the final model. Within each condition, we average $L_P$ over the seeds and report the relative regret $100\,(L_P-L^\star)/L^\star$, where $L^\star$ is the smallest seed-averaged loss among the four methods. Because this regret is bounded below and skewed, we report both its mean and its median. 

\paragraph{Census experiment.}
For the real-data experiment, we use the Adult census data~\cite{becker1996adult}. Its training split of $32{,}561$ people defines the target distribution $p$, and its test split of $16{,}281$ people is used only to score the income questions. Each person is encoded in $n=10$ bits, two Gray-coded bits each for age, education and weekly working hours, and one bit each for marital status, sex, a managerial or professional occupation, and income above 50K. The encoding was fixed before any training, and $845$ of its $1024$ strings occur in the data. The Born machine has $6$ strongly entangling layers with $180$ parameters, so $P_{\mathrm{circ}}=361$, and its $L=120$ noisy locations have the mean error rate of $1.6\%$. The kernel rule, the white-noise simulation, the $T=500$ Adam updates and the controller settings follow the synthetic study, with a learning rate of $\max\{0.005,\,0.1\times0.5^{t/150}\}$ at step $t$. We draw $m\in\{64,128,256,512\}$ people and hold out $20\%$ of them. The unit budget is $U=T\,P_{\mathrm{circ}}\times2000\approx3.6\times10^8$ executions, and $B\in\{1,5,25,125,625\}\times U$. All scores are computed from the exact output distribution $q$ of the final model at the noise level at which it is deployed. The census shares compare the probability that $q$ assigns to an attribute with its share in $p$.

\subsection*{Use of AI tools}
AI tools were used to assist with language polishing and surveying related work.

\section*{Data availability}
Data generated and analyzed during the current study are available from the corresponding author upon reasonable request.
 
\section*{Code availability}
Code used for the current study is available from the
corresponding author upon reasonable request.
 
\section*{Author contributions}
H.M.N. performed the theoretical analysis, developed the methods, and designed the experiments. Q.N. implemented and carried out the experiments. J.T.S. and W.X. contributed to the conceptual development of the work. M.T.T. supervised the project. All authors discussed the results and contributed to the manuscript.
 
\section*{Competing interests}
The authors declare no competing interests.

\bibliography{bibliography}
\newpage

\clearpage
\section*{Supplementary Information}
\setcounter{equation}{0}
\renewcommand{\theequation}{S\arabic{equation}}
\newcounter{savedeq}
\newcommand{\restateresult}[1]{\setcounter{savedeq}{\value{equation}}#1\setcounter{equation}{\value{savedeq}}}

These Supplementary Notes give complete proofs of the results stated in the main text. Supplementary Note~1 collects the Walsh--Fourier facts that underlie the spectral analysis. Supplementary Note~2 proves the generalization bounds, Supplementary Note~3 proves the results on capped error reduction, and Supplementary Note~4 proves the results on budget-aware noise control. Each result is restated with its number from the main text before its proof.

\subsection*{Supplementary Note 1. Walsh--Fourier coordinates}
\label{subsec:prelim-walsh}

The coordinates in which quantum noise acts most simply are the characters of
$\mathbb{Z}_2^n$. For $A\subseteq[n]$ define the Walsh character and, for a
distribution $q$ and a function $\phi$ on $\X$, the associated coefficients
\begin{align}
\chi_A(x):=(-1)^{\sum_{i\in A}x_i},
\qquad
\widehat q(A):=\E_{x\sim q}\bigl[\chi_A(x)\bigr],
\qquad
\widehat\phi(A):=2^{-n}\sum_{t\in\X}\phi(t)\chi_A(t).
\label{eq:prelim-walsh}
\end{align}
Characters are multiplicative, $\chi_A(x\oplus y)=\chi_A(x)\chi_A(y)$, and
orthonormal under the uniform measure. Three facts will be used.

\paragraph{(i) Coefficients are Pauli observables.}
Since $Z_A\ket{x}=\chi_A(x)\ket{x}$ for
$Z_A:=\bigotimes_{i\in A}Z_i$,
\begin{align}
\widehat{q_\theta^{\mathcal{N}}}(A)
=
\Tr\bigl[\rho_\theta^{\mathcal{N}}Z_A\bigr].
\label{eq:prelim-moment-pauli}
\end{align}
The Walsh coefficients of the output distribution are exactly the
computational-basis Pauli correlators of the output state, so $|A|$ is the
order of the correlation described by the mode.

\paragraph{(ii) Parseval links the spectrum to the collision probability.}
Writing $\mathrm{coll}(q):=\sum_{x}q(x)^2$ for the probability that two
independent draws from $q$ agree,
\begin{align}
\sum_{A\subseteq[n]}\widehat q(A)^2
=
2^n\,\mathrm{coll}(q),
\qquad
\sum_{A\subseteq[n]}\widehat\phi(A)^2
=
\E_{x,y\sim u}\bigl[\phi(x\oplus y)^2\bigr].
\label{eq:prelim-parseval}
\end{align}
The uniform distribution attains $\mathrm{coll}(u)=2^{-n}$, so
$2^n\mathrm{coll}(q)$ measures how far $q$ is from flat and, by
equation~\eqref{eq:prelim-parseval}, simultaneously budgets the total
spectral mass available to the generator. This is the quantity through which
the ansatz enters our complexity bound.

\paragraph{(iii) Translation-invariant kernels are diagonal.}
We call $\kappa$ translation invariant if
\begin{align}
\kappa(x,y)=\phi(x\oplus y)
\qquad\text{for some }\phi:\X\to\mathbb{R},
\quad \phi(\mathbf 0)=1,
\label{eq:prelim-ti-kernel}
\end{align}
the normalization being without loss of generality. The Gram matrix $G$ with entries $G_{xy}=\kappa(x,y)$ is then
diagonalized by the characters, with $G\chi_A=2^n\widehat\phi(A)\chi_A$, so
$\kappa$ is positive semidefinite if and only if $\widehat\phi(A)\ge0$ for
all $A$, and in that case $\sum_A\widehat\phi(A)=\phi(\mathbf 0)=1$. The
Gaussian kernel on the Hamming distance used in the main text,
$\kappa(x,y)=\frac1c\sum_{i=1}^c\exp[-d_H(x,y)/(2\tau_i^2)]$, is of this form with
$\phi(t)=\frac1c\sum_{i=1}^ca_i^{|t|}$ and $a_i=e^{-1/(2\tau_i^2)}$.

Under equation~\eqref{eq:prelim-ti-kernel}, multiplicativity of the
characters gives the witness its spectral form,
\begin{align}
h_\theta^{\mathcal{N}}(x)
=
\sum_{A\subseteq[n]}
\widehat\phi(A)\,\widehat{q_\theta^{\mathcal{N}}}(A)\,\chi_A(x),
\label{eq:prelim-witness-fourier}
\end{align}
so the witness is a kernel-weighted superposition of the generator's Pauli
correlators. In particular the uniform distribution has
$\widehat u(A)=0$ for $A\neq\varnothing$, so its witness
$h_u\equiv\widehat\phi(\varnothing)$ is constant. By
equation~\eqref{eq:prelim-centering}, the class may therefore be recentered at no cost,
\begin{align}
(P_m-P)h_\theta^{\mathcal{N}}
=
(P_m-P)\bar h_\theta^{\mathcal{N}},
\qquad
\bar h_\theta^{\mathcal{N}}
:=
h_\theta^{\mathcal{N}}-h_u
=
\sum_{A\neq\varnothing}
\widehat\phi(A)\widehat{q_\theta^{\mathcal{N}}}(A)\chi_A .
\label{eq:prelim-centered-witness}
\end{align}
We write $\bar{\mathcal{H}}^{\mathcal{N}}$ and
$\bar{\mathcal{H}}_\pm^{\mathcal{N}}$ for the corresponding centered and
symmetrized classes, and $\bar h_\theta:=\bar h^0_\theta$ in the noiseless case.

\subsection*{Supplementary Note 2. Proofs of the generalization bounds}

We first prove the reduction to the learned witness, which underlies both generalization theorems, and then the noise-agnostic bound, the noiseless complexity bound, and the bounds under scrambled local Pauli noise.

\restateresult{\lemmaboundtowitness*}

\begin{proof}[Proof of Lemma~\ref{lem:expected-gap-identity}]
For brevity, we condition on $\Ss$ and $\hat\theta_{\mathcal N}$ and write
\begin{align}
q
:=
q_{\hat\theta_{\mathcal N}}^{\mathcal N},
\qquad
h
:=
h_{\hat\theta_{\mathcal N}}^{\mathcal N}.
\end{align}
Under this conditioning, the learned distribution $q$ and the witness $h$
are fixed, while
\begin{align}
\Y=(y_1,\ldots,y_N)\sim q^{\otimes N}.
\end{align}

We set the population target-only and generator-only terms by
\begin{align}
A_p
&:=
\E_{x,x'\sim p}\left[\kappa(x,x')\right],
&
A_q
&:=
\E_{y,y'\sim q}\left[\kappa(y,y')\right].
\end{align}
Since $Ph=\E_{x\sim p,\,y\sim q}[\kappa(x,y)]$ by the definition of the witness,
equation~\eqref{eq:prelim-witness}, the population risk can then be written as
\begin{align}
R_{\mathcal N}(\hat\theta_{\mathcal N})
=
A_q+A_p-2Ph.
\label{eq:proof-population-decomposition}
\end{align}

Similarly, we write the empirical risk as
\begin{align}
\widehat R_{\Ss,\Y}^{\mathcal N}(\hat\theta_{\mathcal N})
=
U_{\Y}+U_{\Ss}-2\widehat C,
\label{eq:proof-empirical-decomposition}
\end{align}
where
\begin{align}
U_{\Ss}
&:=
\frac{1}{m(m-1)}
\sum_{i\neq i'}
\kappa(x_i,x_{i'}),
\\
U_{\Y}
&:=
\frac{1}{N(N-1)}
\sum_{j\neq j'}
\kappa(y_j,y_{j'}),
\\
\widehat C
&:=
\frac{1}{mN}
\sum_{i=1}^{m}\sum_{j=1}^{N}
\kappa(x_i,y_j).
\end{align}

Because the evaluation samples are conditionally i.i.d.\ from $q$, every
pair $(y_j,y_{j'})$ with $j\neq j'$ has joint distribution $q\otimes q$.
Therefore,
\begin{align}
\E_{\Y}
\left[
U_{\Y}
\,\middle|\,
\Ss,\hat\theta_{\mathcal N}
\right]
&=
\frac{1}{N(N-1)}
\sum_{j\neq j'}
\E_{y_j,y_{j'}\sim q}
\left[\kappa(y_j,y_{j'})\right]
\\
&=
\frac{1}{N(N-1)}
\sum_{j\neq j'} A_q
=
A_q.
\label{eq:proof-generator-unbiased}
\end{align}

For the empirical coupling term, each $x_i$ is fixed under the conditioning,
while every $y_j$ is distributed according to $q$. Hence,
\begin{align}
\E_{\Y}
\left[
\widehat C
\,\middle|\,
\Ss,\hat\theta_{\mathcal N}
\right]
&=
\frac{1}{mN}
\sum_{i=1}^{m}\sum_{j=1}^{N}
\E_{y_j\sim q}
\left[\kappa(x_i,y_j)\right]
\\
&=
\frac{1}{mN}
\sum_{i=1}^{m}\sum_{j=1}^{N}
h(x_i)
\\
&=
\frac{1}{m}\sum_{i=1}^{m}h(x_i)
=
P_mh.
\label{eq:proof-coupling-expectation}
\end{align}

Substituting equations~\eqref{eq:proof-generator-unbiased} and
\eqref{eq:proof-coupling-expectation} into
equations~\eqref{eq:proof-population-decomposition} and
\eqref{eq:proof-empirical-decomposition} yields
\begin{align}
&\E_{\Y}
\left[
R_{\mathcal N}(\hat\theta_{\mathcal N})
-
\widehat R_{\Ss,\Y}^{\mathcal N}(\hat\theta_{\mathcal N})
\,\middle|\,
\Ss,\hat\theta_{\mathcal N}
\right]
\nonumber\\
&\quad=
A_q+A_p-2Ph
-
\left(A_q+U_{\Ss}-2P_mh\right)
\\
&\quad=
A_p-U_{\Ss}
+
2(P_m-P)h.
\label{eq:proof-conditional-gap}
\end{align}

It remains to average over the target sample and the learning algorithm.
Since $x_i$ and $x_{i'}$ are independent draws from $p$ whenever
$i\neq i'$,
\begin{align}
\E_{\Ss}[U_{\Ss}]
&=
\frac{1}{m(m-1)}
\sum_{i\neq i'}
\E_{x_i,x_{i'}\sim p}
\left[\kappa(x_i,x_{i'})\right]
\\
&=
\frac{1}{m(m-1)}
\sum_{i\neq i'}A_p
=
A_p.
\end{align}
Thus, $\E_{\Ss}[A_p-U_{\Ss}]=0$. Taking the remaining expectations in
equation~\eqref{eq:proof-conditional-gap} gives
\begin{align}
\Delta_{\mathcal N}(m)
=
2\,
\E_{\Ss,\A_{\mathcal N}}
\left[
(P_m-P)h_{\hat\theta_{\mathcal N}}^{\mathcal N}
\right],
\end{align}
which proves the claim.
\end{proof}

\restateresult{\theoremnoiseagnostic*}

\begin{proof}[Proof of Theorem~\ref{thm:noise-agnostic-generalization}]
Using Lemma~\ref{lem:expected-gap-identity},
\begin{align}
\left|\Delta_{\mathcal N}(m)\right|
&\le
2\,
\E_{\Ss,\A_{\mathcal N}}
\left[
\left|
(P_m-P)h_{\hat\theta_{\mathcal N}}^{\mathcal N}
\right|
\right]
\\
&\le
2\,
\E_{\Ss}
\left[
\sup_{h\in\W^{\mathcal N}}
\left|(P_m-P)h\right|
\right]
\\
&\le
4\,
\Rad\left(\W_\pm^{\mathcal N}\right),
\end{align}
where the last step is the symmetrization bound in
equation~\eqref{eq:uniform-deviation}.
\end{proof}

\restateresult{\lemmanoiseless*}

\begin{proof}[Proof of Lemma~\ref{lem:rademacher-anticoncentration}]
We first translate the concentration condition into a bound on the
nonconstant Fourier moments of the generator. Under the convention in
equation~\eqref{eq:fourier-conventions}, Parseval's identity gives
\begin{align}
\sum_{A\subseteq[n]}
\widehat{q_\theta^0}(A)^2
=
2^n\sum_{x\in\X}q_\theta^0(x)^2.
\label{eq:parseval-generator}
\end{align}
The empty coefficient satisfies
\begin{align}
\widehat{q_\theta^0}(\varnothing)
=
\E_{x\sim q_\theta^0}[\chi_\varnothing(x)]
=
1.
\end{align}
Therefore, removing the empty mode from
equation~\eqref{eq:parseval-generator}, we obtain
\begin{align}
\sum_{A\neq\varnothing}
\widehat{q_\theta^0}(A)^2
&=
2^n\sum_{x\in\X}q_\theta^0(x)^2-1
\notag\\
&\leq
C-1
\qquad
\text{for every $\theta\in\Theta$}.
\label{eq:spectral-budget}
\end{align}
Thus, centering removes the constant Fourier coefficient and changes the
available spectral budget from $C$ to $C-1$.

Now fix an arbitrary training sample
$\Ss=(x_1,\ldots,x_m)$ and let
$\sigma_1,\ldots,\sigma_m$ be independent Rademacher variables. For a class $\mathcal F$, we write
\begin{align}
\widehat{\Rad}_{\Ss}(\mathcal F):=\E_\sigma\Bigl[\sup_{f\in\mathcal F}\frac1m\sum_{i=1}^m\sigma_if(x_i)\Bigr]
\label{eq:empirical-rademacher}
\end{align}
for its empirical Rademacher complexity, so that $\Rad(\mathcal F)=\E_{\Ss}[\widehat{\Rad}_{\Ss}(\mathcal F)]$ by equation~\eqref{eq:prelim-rademacher}.
For each nonempty $A\subseteq[n]$, we define
\begin{align}
g_A
:=
\frac{1}{m}
\sum_{i=1}^m
\sigma_i\chi_A(x_i).
\end{align}
Using the spectral representation of the centered witness in equation~\eqref{eq:witness-fourier}, we obtain
\begin{align}
\frac{1}{m}
\sum_{i=1}^m
\sigma_i\bar h_\theta(x_i)
&=
\sum_{A\neq\varnothing}
\widehat{\phi}(A)
\widehat{q_\theta^0}(A)
g_A.
\label{eq:rademacher-spectral-sum}
\end{align}
Because
$\bar{\W}_\pm^0
=\bar{\W}^0\cup(-\bar{\W}^0)$,
its empirical Rademacher complexity is
\begin{align}
\widehat{\Rad}_{\Ss}
\bigl(\bar{\W}_\pm^0\bigr)
&=
\E_\sigma
\left[
\sup_{\theta\in\Theta}
\left|
\sum_{A\neq\varnothing}
\widehat{\phi}(A)
\widehat{q_\theta^0}(A)
g_A
\right|
\right].
\label{eq:empirical-rademacher-spectral}
\end{align}

For every $\theta$, based on Cauchy--Schwarz and
equation~\eqref{eq:spectral-budget}, we obtain
\begin{align}
\left|
\sum_{A\neq\varnothing}
\widehat{\phi}(A)
\widehat{q_\theta^0}(A)
g_A
\right|
&\leq
\left(
\sum_{A\neq\varnothing}
\widehat{q_\theta^0}(A)^2
\right)^{1/2}
\left(
\sum_{A\neq\varnothing}
\widehat{\phi}(A)^2g_A^2
\right)^{1/2}
\notag\\
&\leq
\sqrt{C-1}
\left(
\sum_{A\neq\varnothing}
\widehat{\phi}(A)^2g_A^2
\right)^{1/2}.
\end{align}
The final expression is independent of $\theta$. Hence, by Jensen's
inequality,
\begin{align}
\widehat{\Rad}_{\Ss}
\bigl(\bar{\W}_\pm^0\bigr)
&\leq
\sqrt{C-1}
\left(
\sum_{A\neq\varnothing}
\widehat{\phi}(A)^2
\E_\sigma[g_A^2]
\right)^{1/2}.
\label{eq:jensen-rademacher}
\end{align}

It remains to evaluate $\E_\sigma[g_A^2]$. Independence and centering of
the Rademacher variables eliminate all cross terms, while
$\chi_A(x_i)^2=1$. Consequently,
\begin{align}
\E_\sigma[g_A^2]
&=
\frac{1}{m^2}
\sum_{i,i'=1}^m
\E_\sigma[\sigma_i\sigma_{i'}]
\chi_A(x_i)\chi_A(x_{i'})
\notag\\
&=
\frac{1}{m^2}
\sum_{i=1}^m
\chi_A(x_i)^2
=
\frac{1}{m}.
\label{eq:rademacher-mode-variance}
\end{align}
Substituting this identity into
equation~\eqref{eq:jensen-rademacher} gives, for every fixed sample $\Ss$,
\begin{align}
\widehat{\Rad}_{\Ss}
\bigl(\bar{\W}_\pm^0\bigr)
&\leq
\sqrt{
\frac{
(C-1)
\displaystyle\sum_{A\neq\varnothing}\widehat{\phi}(A)^2
}{m}
}.
\end{align}
For the second branch of $\Lambda_C$, we apply the Cauchy--Schwarz inequality with the weights $\widehat\phi(A)\ge0$ instead such that:
\begin{align}
\left|\sum_{A\neq\varnothing}\widehat{\phi}(A)\widehat{q_\theta^0}(A)g_A\right|
&\leq
\Bigl(\sum_{A\neq\varnothing}\widehat{\phi}(A)\widehat{q_\theta^0}(A)^2\Bigr)^{1/2}
\Bigl(\sum_{A\neq\varnothing}\widehat{\phi}(A)g_A^2\Bigr)^{1/2}\\
&\leq
\bigl(1-\widehat\phi(\varnothing)\bigr)^{1/2}
\Bigl(\sum_{A\neq\varnothing}\widehat{\phi}(A)g_A^2\Bigr)^{1/2},
\end{align}
where the last step uses $|\widehat{q_\theta^0}(A)|\le1$ and $\sum_{A}\widehat\phi(A)=\phi(\mathbf 0)=1$. The right-hand side does not depend on $\theta$, so Jensen's inequality and equation~\eqref{eq:rademacher-mode-variance} give:
\begin{equation}
\widehat{\Rad}_{\Ss}\bigl(\bar{\W}_\pm^0\bigr)
\leq
\bigl(1-\widehat\phi(\varnothing)\bigr)^{1/2}\Bigl(\frac1m\sum_{A\neq\varnothing}\widehat{\phi}(A)\Bigr)^{1/2}
=\frac{1-\widehat\phi(\varnothing)}{\sqrt m}.
\end{equation}
Both bounds hold for every fixed sample, and so does their minimum $\Lambda_C/\sqrt m$. Taking expectation over $\Ss$ proves
equation~\eqref{eq:population-rademacher-bound}.
\end{proof}

\restateresult{\theorempauli*}

\restateresult{\corollarypauli*}

\begin{proof}[Proof of Theorem~\ref{thm:effective-depolarizing-generalization} and Corollary~\ref{cor:effective-depolarizing-high-probability}]
Let
\begin{align}
\mu_{P_m}
&:=
\frac{1}{m}
\sum_{i=1}^m
\kappa(x_i,\cdot),
&
\mu_p
&:=
\E_{x\sim p}
\bigl[
\kappa(x,\cdot)
\bigr]
\end{align}
denote the empirical and population kernel mean embeddings of the target
distribution. By the reproducing property, for every $\theta\in\Theta$,
\begin{align}
(P_m-P)e_\theta
=
\left\langle
e_\theta,
\mu_{P_m}-\mu_p
\right\rangle_{\Hil}.
\end{align}
Cauchy--Schwarz and
equation~\eqref{eq:witness-effective-error-bound} therefore give
\begin{align}
\sup_{\theta\in\Theta}
\left|
(P_m-P)e_\theta
\right|
\leq
\varepsilon_{\mathrm{Pauli}}
\left\|
\mu_{P_m}-\mu_p
\right\|_{\Hil}.
\label{eq:residual-mean-embedding-bound}
\end{align}

Because $\kappa(x,x)=1$, independence of the target samples gives
\begin{align}
\E_{\Ss}
\left[
\left\|
\mu_{P_m}-\mu_p
\right\|_{\Hil}^2
\right]
&=
\frac{1}{m}
\E_{x\sim p}
\left[
\left\|
\kappa(x,\cdot)-\mu_p
\right\|_{\Hil}^2
\right]
\notag\\
&=
\frac{1}{m}
\left(
\E_{x\sim p}[\kappa(x,x)]
-
\|\mu_p\|_{\Hil}^2
\right)
\leq
\frac{1}{m}.
\label{eq:target-mean-embedding-second-moment}
\end{align}
Jensen's inequality then implies
\begin{align}
\E_{\Ss}
\left[
\left\|
\mu_{P_m}-\mu_p
\right\|_{\Hil}
\right]
\leq
\frac{1}{\sqrt m}.
\label{eq:target-mean-embedding-expectation}
\end{align}

\paragraph{Expected bound (Theorem~\ref{thm:effective-depolarizing-generalization}).}
By Lemma~\ref{lem:expected-gap-identity},
\begin{align}
\Delta_{\mathcal N}(m)
=
2\E_{\Ss,\A_{\mathcal N}}
\left[
(P_m-P)
h_{\hat\theta_{\mathcal N}}^{\mathcal N}
\right].
\end{align}
Using equation~\eqref{eq:physical-effective-witness-decomposition} and the
fact that $P_m-P$ annihilates constants, we obtain
\begin{align}
\left|\Delta_{\mathcal N}(m)\right|
&\leq
2(1-\gamma)
\E_{\Ss}
\left[
\sup_{\theta\in\Theta}
\left|
(P_m-P)\bar h_\theta
\right|
\right]
\notag\\
&\quad
+
2\E_{\Ss}
\left[
\sup_{\theta\in\Theta}
\left|
(P_m-P)e_\theta
\right|
\right].
\label{eq:effective-expected-reduction}
\end{align}
Standard symmetrization gives
\begin{align}
\E_{\Ss}
\left[
\sup_{\theta\in\Theta}
\left|
(P_m-P)\bar h_\theta
\right|
\right]
\leq
2\Rad
\bigl(
\bar{\W}_\pm^0
\bigr).
\end{align}
equations~\eqref{eq:residual-mean-embedding-bound} and
\eqref{eq:target-mean-embedding-expectation} give
\begin{align}
\E_{\Ss}
\left[
\sup_{\theta\in\Theta}
\left|
(P_m-P)e_\theta
\right|
\right]
\leq
\frac{\varepsilon_{\mathrm{Pauli}}}{\sqrt m}.
\end{align}
Applying Lemma~\ref{lem:rademacher-anticoncentration} proves
equation~\eqref{eq:effective-depolarizing-expected-bound}.

\paragraph{High-probability bound (Corollary~\ref{cor:effective-depolarizing-high-probability}).}
Define the target-only U-statistic
\begin{align}
U_{\Ss}
:=
\frac{1}{m(m-1)}
\sum_{i\neq i'}
\kappa(x_i,x_{i'})
\end{align}
and, for fixed $\Ss$ and $\theta$, collect the generator-dependent terms in
\begin{align}
G_{\Ss,\theta}(\Y)
:={}&
\frac{1}{N(N-1)}
\sum_{j\neq j'}
\kappa(y_j,y_{j'})
\notag\\
&-
\frac{2}{mN}
\sum_{i=1}^m
\sum_{j=1}^N
\kappa(x_i,y_j).
\end{align}
Conditioning on $\Ss$ and $\hat\theta_{\mathcal N}$, the fresh-sample
assumption gives
\begin{align}
&
R_{\mathcal N}(\hat\theta_{\mathcal N})
-
\widehat R_{\Ss,\Y}^{\mathcal N}
(\hat\theta_{\mathcal N})
\notag\\
={}&
\underbrace{
\E_{x,x'\sim p}
[\kappa(x,x')]
-
U_{\Ss}
}_{\text{target U-statistic fluctuation}}
+
2(P_m-P)
h_{\hat\theta_{\mathcal N}}^{\mathcal N}
\notag\\
&+
\underbrace{
\E_{\Y}
\left[
G_{\Ss,\hat\theta_{\mathcal N}}(\Y)
\,\middle|\,
\Ss,\hat\theta_{\mathcal N}
\right]
-
G_{\Ss,\hat\theta_{\mathcal N}}(\Y)
}_{\text{generator-sample fluctuation}}.
\label{eq:effective-high-probability-decomposition}
\end{align}

We first control the noisy witness term. From
equation~\eqref{eq:physical-effective-witness-decomposition},
\begin{align}
2\left|
(P_m-P)
h_{\hat\theta_{\mathcal N}}^{\mathcal N}
\right|
&\leq
2(1-\gamma)
\sup_{\theta\in\Theta}
\left|
(P_m-P)\bar h_\theta
\right|
\notag\\
&\quad
+
2\varepsilon_{\mathrm{Pauli}}
\left\|
\mu_{P_m}-\mu_p
\right\|_{\Hil}.
\label{eq:noisy-witness-high-probability-reduction}
\end{align}

Since $0\leq h_\theta^0\leq1$ and
$\widehat{\phi}(\varnothing)\in[0,1]$, a standard Rademacher deviation
inequality gives, with probability at least $1-\beta/4$,
\begin{align}
\sup_{\theta\in\Theta}
\left|
(P_m-P)\bar h_\theta
\right|
\leq
2\Rad
\bigl(
\bar{\W}_\pm^0
\bigr)
+
\sqrt{
\frac{2\log(4/\beta)}{m}
}.
\label{eq:centered-rademacher-high-probability}
\end{align}

Changing one target sample changes
$\|\mu_{P_m}-\mu_p\|_{\Hil}$ by at most $2/m$. Therefore,
McDiarmid's inequality and
equation~\eqref{eq:target-mean-embedding-expectation} imply that, with
probability at least $1-\beta/4$,
\begin{align}
\left\|
\mu_{P_m}-\mu_p
\right\|_{\Hil}
\leq
\frac{1}{\sqrt m}
+
\sqrt{
\frac{2\log(4/\beta)}{m}
}.
\label{eq:target-mean-embedding-concentration}
\end{align}
Combining these two inequalities with
Lemma~\ref{lem:rademacher-anticoncentration} bounds the noisy witness term by
\begin{align}
&
4(1-\gamma)
\frac{\Lambda_C}{\sqrt m}
+
2(1-\gamma)
\sqrt{
\frac{2\log(4/\beta)}{m}
}
\notag\\
&\qquad
+
\frac{2\varepsilon_{\mathrm{Pauli}}}{\sqrt m}
\left(
1+\sqrt{2\log(4/\beta)}
\right).
\label{eq:noisy-witness-high-probability-bound}
\end{align}

Next, Hoeffding's inequality for order-two U-statistics gives, with
probability at least $1-\beta/4$,
\begin{align}
\left|
\E_{x,x'\sim p}[\kappa(x,x')]
-
U_{\Ss}
\right|
\leq
\sqrt{
\frac{\log(8/\beta)}
{2\lfloor m/2\rfloor}
}.
\label{eq:target-u-statistic-concentration}
\end{align}

Finally, changing one generator sample changes
$G_{\Ss,\theta}(\Y)$ by at most $4/N$. Conditional McDiarmid's inequality
therefore gives, with conditional probability at least $1-\beta/4$,
\begin{align}
&
\left|
\E_{\Y}
\left[
G_{\Ss,\hat\theta_{\mathcal N}}(\Y)
\,\middle|\,
\Ss,\hat\theta_{\mathcal N}
\right]
-
G_{\Ss,\hat\theta_{\mathcal N}}(\Y)
\right|
\leq
\sqrt{
\frac{8\log(8/\beta)}{N}
}.
\label{eq:generator-sample-concentration}
\end{align}
Combining these four events and applying a union bound proves
equation~\eqref{eq:effective-depolarizing-high-probability}.
\end{proof}

\subsection*{Supplementary Note 3. Proofs for capped error reduction}

This note proves the cost of a mitigation profile and the optimality of the capped allocation.

\restateresult{\lemmaprofilecost*}

\begin{proof}[Proof of Lemma~\ref{lem:profile-cost}]
Let
\begin{align}
\delta_k
:=
\epsilon_k-r_k
\end{align}
denote the amount of noise removed at location $k$. For a Pauli operator $Q$,
let $a_k(Q)$ be the probability that an error drawn from $\pi_k$
anticommutes with $Q$. The Pauli-transfer eigenvalues of the physical and
residual channels are, respectively,
\begin{align}
1-2\epsilon_k a_k(Q)
\qquad\text{and}\qquad
1-2r_k a_k(Q).
\end{align}
Since $\epsilon_k<1/2$, the physical channel is invertible. Hence, the
correction map
$\mathcal M_{k,r_k}
=\mathcal E_{k,r_k}\circ\mathcal E_k^{-1}$
has eigenvalues
\begin{align}
\frac{1-2r_k a_k(Q)}
{1-2\epsilon_k a_k(Q)}
=
1+2\delta_k a_k(Q)
+
\mathcal O(\epsilon_k\delta_k).
\end{align}
Transforming these eigenvalues back to the Pauli quasiprobability
representation gives
\begin{align}
\eta_k(I)
&=
1+\delta_k
+
\mathcal O(\epsilon_k\delta_k),
\\
\eta_k(P)
&=
-\delta_k\pi_k(P)
+
\mathcal O(\epsilon_k\delta_k),
\qquad
P\neq I.
\end{align}
Using $\sum_{P\neq I}\pi_k(P)=1$, the local normalization therefore satisfies
\begin{align}
\Gamma_k(r_k)
:=
\sum_P|\eta_k(P)|
=
1+2\delta_k
+
\mathcal O(\epsilon_k\delta_k).
\label{eq:local-profile-cost}
\end{align}

The quasiprobability normalizations multiply across circuit locations, so
\begin{align}
\log\Gamma(\mathbf r)
&=
\sum_{k=1}^L\log\Gamma_k(r_k)
\notag\\
&=
\sum_{k=1}^L
\log\left[
1+2\delta_k
+
\mathcal O(\epsilon_k\delta_k)
\right]
\notag\\
&=
2\sum_{k=1}^L\delta_k
+
\mathcal O\left(
\sum_{k=1}^L\epsilon_k\delta_k
\right),
\end{align}
where we used
$\log(1+x)=x+\mathcal O(x^2)$ and
$\delta_k^2\leq\epsilon_k\delta_k$. Finally,
\begin{align}
\sum_{k=1}^L\delta_k
=
\sum_{k=1}^L(\epsilon_k-r_k)
=
\norm{\boldsymbol\epsilon-\mathbf r}_1,
\end{align}
which proves equation~\eqref{eq:profile-cost}. Neglecting the accumulated
higher-order term in the weak-noise regime and exponentiating yields
equation~\eqref{eq:leading-profile-cost}.
\end{proof}

\restateresult{\lemmacappedprofile*}

\begin{proof}[Proof of Lemma~\ref{lem:capped-profile}]
It is equivalent to minimize $\sum_k r_k^2$. Suppose that two residual rates
satisfy $r_i<r_j$ and $r_i<\epsilon_i$. Moving a sufficiently small amount
$\delta>0$ from $r_j$ to $r_i$ preserves the total residual noise and all
feasibility constraints, while changing the objective by
\begin{align}
&(r_i+\delta)^2+(r_j-\delta)^2-r_i^2-r_j^2
\notag\\
&\qquad=
2\delta(r_i-r_j)+2\delta^2
<0.
\end{align}
Thus, all residual rates not limited by their physical upper bounds must be
equal. Locations whose physical rates are below this common level remain
unchanged. Hence,
\begin{align}
r_k^{\mathrm{cap}}(R)
=
\min\{\epsilon_k,\nu(R)\}.
\end{align}
The budget constraint determines $\nu(R)$, and uniqueness follows from the
strict convexity of $\sum_k r_k^2$.
\end{proof}

\subsection*{Supplementary Note 4. Proofs for budget-aware noise control}

This note proves the sensitivity of the risk to parameter changes, the cumulative regret bound and its rate.

\restateresult{\lemmarisklipschitz*}

\begin{proof}[Proof of Lemma~\ref{lem:risk-lipschitz}]
Fix a residual profile $\mathbf r$ and write $q_\theta:=q^{\mathbf r}_\theta$. Since $R_{\mathbf r}(\theta)=\norm{\mu_{q_\theta}-\mu_p}_{\Hil}^2$, its partial derivatives are
\begin{align}
\partial_kR_{\mathbf r}(\theta)=2\inner{\partial_k\mu_{q_\theta}}{\mu_{q_\theta}-\mu_p}_{\Hil}.
\label{eq:risk-partial}
\end{align}
The parameter $\theta_k$ enters only through the rotation $e^{-i\theta_kP_k/2}$ with $P_k^2=I$, and every other operation in the circuit, including the noise channels and the measurement, is fixed and linear. For any state $\rho$ entering the rotation,
\begin{align}
e^{-i\theta_kP_k/2}\,\rho\, e^{i\theta_kP_k/2}
=\frac{\rho+P_k\rho P_k}{2}
+\cos\theta_k\,\frac{\rho-P_k\rho P_k}{2}
-\frac{i\sin\theta_k}{2}\,[P_k,\rho],
\end{align}
so each probability $q_\theta(x)$ is a trigonometric polynomial of degree one in $\theta_k$. The parameter-shift rule therefore holds exactly, and by linearity of the mean embedding,
\begin{align}
\partial_k\mu_{q_\theta}
=\frac12\Bigl(\mu_{q_{\theta+\frac\pi2e_k}}-\mu_{q_{\theta-\frac\pi2e_k}}\Bigr),
\end{align}
where $e_k$ is the $k$th unit vector. Both shifted circuits produce valid probability distributions, and $\norm{\mu_q}_{\Hil}^2=\E_{y,y'\sim q}[\kappa(y,y')]\le1$ for every distribution $q$, so $\norm{\partial_k\mu_{q_\theta}}_{\Hil}\le1$. Moreover, $\norm{\mu_{q_\theta}-\mu_p}_{\Hil}^2=\MMD^2_\kappa(q_\theta,p)\le2$ because $\kappa\in[0,1]$ and $\kappa(x,x)=1$. The Cauchy--Schwarz inequality in equation~\eqref{eq:risk-partial} then gives $|\partial_kR_{\mathbf r}(\theta)|\le2\sqrt2$ for every $k$, hence $\norm{\nabla R_{\mathbf r}(\theta)}_2\le2\sqrt{2d_\theta}$. Since $R_{\mathbf r}$ is smooth in $\theta$, the mean value theorem along the segment from $\theta'$ to $\theta$ yields equation~\eqref{eq:regret-risk-lipschitz}. None of these bounds depends on $\mathbf r$, so the constant is uniform over residual profiles.
\end{proof}

\restateresult{\theoremregret*}

\begin{proof}[Proof of Theorem~\ref{thm:controller-regret}]
At each comparison step $s\in\mathcal I$, let
$j_s^\star\in\arg\min_{j\in\mathcal J}\Psi_s(j)$.
The selected profile $j_s$ is retained for
$w_s=\min\{K,T-s\}$ steps. Since comparisons start at $s=0$,
these blocks partition the training trajectory. Regrouping the
terms in $\operatorname{Reg}_T$ gives the exact decomposition
\begin{align}
\operatorname{Reg}_T
={}&
\underbrace{
\sum_{s\in\mathcal I}w_s
\left[\Psi_s(j_s)-\Psi_s(j_s^\star)\right]
}_{\text{selection error at comparison steps}}
\notag\\
&+
\underbrace{
\sum_{s\in\mathcal I}
\sum_{t=s}^{s+w_s-1}
\left[
\Psi_t(j_s)-\Psi_s(j_s)
+\Psi_s(j_s^\star)-\min_j\Psi_t(j)
\right]
}_{\text{score changes between comparisons}}.
\label{eq:regret-block-decomposition}
\end{align}

To decompose the selection error, recall that
$\widetilde L_{\mathrm{val},s}^{U}(j)$ is the validation loss
computed from the fresh PER probes using
equation~\eqref{eq:controller-validation-ustat}.
Let $\mathcal F_s$ contain the target data and all information
available before these probes, including $\theta_s$ and the
probe counts. Define the probe-averaged validation loss by
\begin{align}
\overline L_{\mathrm{val},s}^{U}(j)
:=
\mathbb E\!\left[
\widetilde L_{\mathrm{val},s}^{U}(j)
\,\middle|\,\mathcal F_s
\right].
\end{align}
This average removes the fresh probe randomness while retaining
the same validation sample.

We first bound the selection error in the first sum of
equation~\eqref{eq:regret-block-decomposition}.
At comparison $s$, the controller's selection rule ~\eqref{eq:budget-aware-controller-score} gives
\begin{align}
\widetilde L_{\mathrm{val},s}^{U}(j_s)
-\widetilde L_{\mathrm{val},s}^{U}(j_s^\star)
+\lambda
\frac{\Gamma_{j_s}^2-\Gamma_{j_s^\star}^2}
{N_{\mathrm{future},s}}
\leq b_{sj_s^\star}-b_{sj_s}\leq\bar b_s.
\end{align}
where $\bar b_s:=\max_{j\in\mathcal J}b_{sj}$ and the last step uses $b_{sj}\ge0$. Adding and subtracting the measured and probe-averaged losses
therefore separates the selection error into three contributions,
\begin{align}
&\sum_{s\in\mathcal I}w_s
\left[\Psi_s(j_s)-\Psi_s(j_s^\star)\right]
\notag\\
&\quad\leq
\underbrace{
\sum_{s\in\mathcal I}w_s
\Big[
R_{\mathbf r^{(j_s)}}(\theta_s)
-R_{\mathbf r^{(j_s^\star)}}(\theta_s)
-\overline L_{\mathrm{val},s}^{U}(j_s)
+\overline L_{\mathrm{val},s}^{U}(j_s^\star)
\Big]
}_{\text{validation-error contribution}}
\notag\\
&\qquad+
\underbrace{
\sum_{s\in\mathcal I}w_s
\Big[
\overline L_{\mathrm{val},s}^{U}(j_s)
-\widetilde L_{\mathrm{val},s}^{U}(j_s)
-\overline L_{\mathrm{val},s}^{U}(j_s^\star)
+\widetilde L_{\mathrm{val},s}^{U}(j_s^\star)
\Big]
}_{\text{probe-error contribution}}\notag\\
&\qquad{}+\underbrace{\sum_{s\in\mathcal I}w_s\,\bar b_s}_{\text{radius contribution}}.
\label{eq:regret-comparison-decomposition}
\end{align}
The next three paragraphs bound these validation, probe and radius contributions. The final paragraph bounds the second sum in
equation~\eqref{eq:regret-block-decomposition}, which accounts
for score changes between comparisons.

\paragraph{Validation uncertainty.}
We first bound the validation-error contribution to
$\operatorname{Reg}_T$,
\begin{align}
\sum_{s\in\mathcal I}w_s
\Big[
R_{\mathbf r^{(j_s)}}(\theta_s)
-R_{\mathbf r^{(j_s^\star)}}(\theta_s)
-\overline L_{\mathrm{val},s}^{U}(j_s)
+\overline L_{\mathrm{val},s}^{U}(j_s^\star)
\Big].
\label{eq:regret-validation-contribution}
\end{align}
The expression inside the brackets measures how much the validation
sample distorts the comparison of population losses between two candidates.
We first express this difference using the mean-embedding
representation of MMD, then bound the resulting sampling error.

Using the kernel mean embeddings
\begin{align}
\mu_q:=\mathbb E_{X\sim q}[\kappa(X,\cdot)],
\qquad
\widehat\mu_{\mathrm{val}}
:=\frac1v\sum_{i=1}^{v}\kappa(x_i,\cdot),
\end{align}
we express the loss difference directly as
\begin{align}
R_{\mathbf r^{(j)}}(\theta_s)
-\overline L_{\mathrm{val},s}^{U}(j)
=
\underbrace{\|\mu_p\|_{\Hil}^2-U_{\mathrm{val}}}
_{\text{independent of the candidate}}
+
2\left\langle
\widehat\mu_{\mathrm{val}}-\mu_p,\,
\mu_{q_{\theta_s}^{\mathbf r^{(j)}}}
\right\rangle_{\Hil},
\end{align}
where
$U_{\mathrm{val}}
=[v(v-1)]^{-1}\sum_{i\ne i'}\kappa(x_i,x_{i'})$.
This follows by expanding population MMD and averaging the
validation U-statistic over the independent, unbiased PER probes.

The common term cancels when comparing any candidate $j_s$
with $j_s^\star$, giving
\begin{align}
&
\left|
R_{\mathbf r^{(j_s)}}(\theta_s)
-R_{\mathbf r^{(j_s^\star)}}(\theta_s)
-\overline L_{\mathrm{val},s}^{U}(j_s)
+\overline L_{\mathrm{val},s}^{U}(j_s^\star)
\right|
\notag\\
&\quad=
2\left|
\left\langle
\widehat\mu_{\mathrm{val}}-\mu_p,\,
\mu_{q_{\theta_s}^{\mathbf r^{(j_s)}}}
-\mu_{q_{\theta_s}^{\mathbf r^{(j_s^\star)}}}
\right\rangle_{\Hil}
\right|
\label{eq:regret-validation-contrast}\\
&\quad\leq
2\|\widehat\mu_{\mathrm{val}}-\mu_p\|_{\Hil}
\operatorname{MMD}_{\kappa}
\left(q_{\theta_s}^{\mathbf r^{(j_s)}},
      q_{\theta_s}^{\mathbf r^{(j_s^\star)}}\right)
\qquad\text{(Cauchy--Schwarz)}.
\label{eq:regret-validation-two-factors}
\end{align}

For the validation sampling error,
\begin{align}
\mathbb E\|\widehat\mu_{\mathrm{val}}-\mu_p\|_{\Hil}
&\leq
\sqrt{
\mathbb E\|\widehat\mu_{\mathrm{val}}-\mu_p\|_{\Hil}^2
}
&&\text{(Jensen)}
\notag\\
&=
\sqrt{
\frac{
\mathbb E_{X\sim p}\kappa(X,X)-\|\mu_p\|_{\Hil}^2
}{v}
}
\leq\frac1{\sqrt v}
&&\text{(i.i.d., $\kappa(x,x)\leq1$)}.
\end{align}
Changing one observation changes the norm by at most $2/v$.
Hence, with probability at least $1-\beta/2$,
\begin{align}
\|\widehat\mu_{\mathrm{val}}-\mu_p\|_{\Hil}
&\leq
\mathbb E\|\widehat\mu_{\mathrm{val}}-\mu_p\|_{\Hil}
+\sqrt{\frac{2\log(2/\beta)}v}
&&\text{(McDiarmid)}
\notag\\
&\leq
\frac1{\sqrt v}+\sqrt{\frac{2\log(2/\beta)}v}
=a_v(\beta).
\label{eq:regret-validation-event}
\end{align}

For the distance between candidates, recall
$D_s=\operatorname{MMD}_{\kappa}(q_{\theta_s}^{0},u)$.
The mismatch guarantee and effective depolarization yield
\begin{align}
\operatorname{MMD}_{\kappa}
\left(q_{\theta_s}^{\mathbf r^{(j_s)}},
      q_{\theta_s}^{\mathbf r^{(j_s^\star)}}\right)
&\leq
\operatorname{MMD}_{\kappa}
\left(q_{\theta_s}^{\gamma_{j_s}},
      q_{\theta_s}^{\gamma_{j_s^\star}}\right)
+2\varepsilon_{\mathrm{tol}}
&&\text{(triangle inequality)}
\notag\\
&=
|\gamma_{j_s}-\gamma_{j_s^\star}|D_s
+2\varepsilon_{\mathrm{tol}}
&&\text{(effective depolarization)}
\notag\\
&\leq
\delta_\gamma D_s+2\varepsilon_{\mathrm{tol}}.
\label{eq:regret-candidate-separation}
\end{align}

Combining these bounds in
equation~\eqref{eq:regret-validation-two-factors} controls the
comparison of every candidate $j_s$ with $j_s^\star$ on the same
validation event, including at adaptively chosen parameters. Summing over the comparison blocks gives
\begin{align}
&
\sum_{s\in\mathcal I}w_s
\Big[
R_{\mathbf r^{(j_s)}}(\theta_s)
-R_{\mathbf r^{(j_s^\star)}}(\theta_s)
-\overline L_{\mathrm{val},s}^{U}(j_s)
+\overline L_{\mathrm{val},s}^{U}(j_s^\star)
\Big]
\notag\\
&\quad\leq
2a_v(\beta)\sum_{s\in\mathcal I}w_s
\left(\delta_\gamma D_s+2\varepsilon_{\mathrm{tol}}\right)
\notag\\
&\quad=
2a_v(\beta)
\left(
\delta_\gamma\sum_{s\in\mathcal I}w_sD_s
+2T\varepsilon_{\mathrm{tol}}
\right),
\qquad \sum_{s\in\mathcal I}w_s=T.
\label{eq:regret-validation-total}
\end{align}
This bounds the validation contribution
\eqref{eq:regret-validation-contribution} to cumulative regret.

\paragraph{Probe uncertainty and its budget dependence.}
We next bound the probe-error contribution in
equation~\eqref{eq:regret-comparison-decomposition}.
The triangle inequality gives
\begin{align}
&
\sum_{s\in\mathcal I}w_s
\Big[
\overline L_{\mathrm{val},s}^{U}(j_s)
-\widetilde L_{\mathrm{val},s}^{U}(j_s)
-\overline L_{\mathrm{val},s}^{U}(j_s^\star)
+\widetilde L_{\mathrm{val},s}^{U}(j_s^\star)
\Big]
\notag\\
&\quad\leq
2\sum_{s\in\mathcal I}w_s
\max_{j\in\mathcal J}
\left|
\widetilde L_{\mathrm{val},s}^{U}(j)
-\overline L_{\mathrm{val},s}^{U}(j)
\right|.
\label{eq:regret-probe-contribution}
\end{align}
Because $j_s$ is selected using these same probes, we control
the errors simultaneously over all candidates before applying
the bound to $j_s$ and $j_s^\star$.

Conditional on $\mathcal F_s$, separate the measured loss into
its generator--generator and target--generator terms,
\begin{align}
\widetilde L_{\mathrm{val},s}^{U}(j)
&=U_{\mathrm{val}}+\widehat A_{sj}-2\widehat C_{sj},
\notag\\
\widehat A_{sj}
&:=
\frac{1}{M_s(M_s-1)}
\sum_{a\ne b}
W_{sj,a}W_{sj,b}\kappa(Y_{sj,a},Y_{sj,b}),
\notag\\
\widehat C_{sj}
&:=
\frac1{M_s}\sum_{a=1}^{M_s}
W_{sj,a}
\left(\frac1v\sum_{i=1}^{v}\kappa(x_i,Y_{sj,a})\right).
\end{align}
The kernel in $\widehat A_{sj}$ lies in
$[-\Gamma_j^2,\Gamma_j^2]$, and each summand averaged in
$\widehat C_{sj}$ lies in $[-\Gamma_j,\Gamma_j]$.
Thus,
\begin{align}
\left|
\widehat A_{sj}
-\mathbb E[\widehat A_{sj}\mid\mathcal F_s]
\right|
&\leq
\Gamma_j^2
\sqrt{\frac{2\log(8JH/\beta)}{\lfloor M_s/2\rfloor}}
&&\text{(U-statistic Hoeffding)},
\notag\\
\left|
\widehat C_{sj}
-\mathbb E[\widehat C_{sj}\mid\mathcal F_s]
\right|
&\leq
\Gamma_j
\sqrt{\frac{2\log(8JH/\beta)}{M_s}}
&&\text{(Hoeffding)},
\end{align}
each with conditional failure probability at most $\beta/(4JH)$.
Averaging over the history and taking a union bound over the
$2JH$ errors gives simultaneous coverage with probability
at least $1-\beta/2$.

On this event, since $U_{\mathrm{val}}$ is fixed under
the conditioning,
\begin{align}
&
\left|
\widetilde L_{\mathrm{val},s}^{U}(j)
-\overline L_{\mathrm{val},s}^{U}(j)
\right|
\notag\\
&\quad\leq
\left|
\widehat A_{sj}
-\mathbb E[\widehat A_{sj}\mid\mathcal F_s]
\right|
+
2\left|
\widehat C_{sj}
-\mathbb E[\widehat C_{sj}\mid\mathcal F_s]
\right|
\notag\\
&\quad\leq
\Gamma_j^2
\sqrt{\frac{2\log(8JH/\beta)}{\lfloor M_s/2\rfloor}}
+
2\Gamma_j
\sqrt{\frac{2\log(8JH/\beta)}{M_s}}
\notag\\
&\quad\leq
(\sqrt6+2\sqrt2)\overline\Gamma^2
\sqrt{\frac{\log(8JH/\beta)}{M_s}},
\label{eq:regret-probe-event}
\end{align}
where the last inequality uses
$\lfloor M_s/2\rfloor\geq M_s/3$ and
$1\leq\Gamma_j\leq\overline\Gamma$.

The allocation rule makes the remaining average probe allowance
nondecreasing. Feasibility therefore gives
\begin{align}
M_s
\geq M_0
=
\left\lfloor\frac{\alpha B}{JH}\right\rfloor
\geq\frac{\alpha B}{2JH},
\qquad M_0\geq2.
\label{eq:probe-allocation-lower}
\end{align}
Substituting this into equation~\eqref{eq:regret-probe-event}
and summing over blocks yields
\begin{align}
&
2\sum_{s\in\mathcal I}w_s
\max_{j\in\mathcal J}
\left|
\widetilde L_{\mathrm{val},s}^{U}(j)
-\overline L_{\mathrm{val},s}^{U}(j)
\right|
\notag\\
&\quad\leq
(8+4\sqrt3)\overline\Gamma^2
\sqrt{\frac{JH\log(8JH/\beta)}{\alpha B}}
\sum_{s\in\mathcal I}w_s
\notag\\
&\quad\leq
15\overline\Gamma^2T
\sqrt{\frac{JH\log(8JH/\beta)}{\alpha B}},
\qquad
\sum_{s\in\mathcal I}w_s=T.
\label{eq:regret-probe-budget-bound}
\end{align}
This bounds the cumulative probe contribution
\eqref{eq:regret-probe-contribution}.

\paragraph{Measurement radius.}
The standard error in the radius is computed from the per-shot projection terms
\begin{equation}\label{eq:psi}
\psi_{sj,a} := 2W_{sj,a}\Bigl(\frac{1}{M_s-1}\sum_{b\neq a} W_{sj,b}\,\kappa(Y_{sj,a},Y_{sj,b}) - \frac{1}{v}\sum_{i=1}^{v}\kappa(x_i,Y_{sj,a})\Bigr)
\end{equation}
as
\begin{equation}\label{eq:se-formula}
\widehat{\mathrm{SE}}\bigl(\widetilde{L}^{U}_{\mathrm{val},s}(j)\bigr) := \Bigl[\frac{1}{M_s(M_s-1)}\sum_{a=1}^{M_s}(\psi_{sj,a}-\bar{\psi}_{sj})^{2}\Bigr]^{1/2},
\end{equation}
where $\bar{\psi}_{sj}$ is the mean of the $\psi_{sj,a}$. Since $|W_{sj,a}| \leq \Gamma_j$ and $\kappa \in [0,1]$, every term satisfies $|\psi_{sj,a}| \leq 2\Gamma_j(\Gamma_j+1) \leq 4\Gamma_j^{2}$. The sample variance of such terms is at most $16\Gamma_j^{4} M_s/(M_s-1)$, so $\widehat{\mathrm{SE}} \leq 4\Gamma_j^{2}/\sqrt{M_s-1} \leq 6\Gamma_j^{2}/\sqrt{M_s}$ for $M_s \geq 2$, and thus $\bar{b}_s \leq 6z\,\overline{\Gamma}^{2}/\sqrt{M_s}$. Equation~\eqref{eq:probe-allocation-lower} gives $M_s \geq \alpha B/(2JH)$, and $\log(8JH/\beta) \geq \log 8 > 2$ because $J, H \geq 1$ and $\beta < 1$. As a result, we have:
\begin{equation}\label{eq:radius-contribution}
\sum_{s\in\mathcal{I}} w_s\,\bar{b}_s
\leq 6z\,\overline{\Gamma}^{2}\,T\sqrt{\frac{2JH}{\alpha B}}
\leq 6z\,\overline{\Gamma}^{2}\,T\sqrt{\frac{JH\log(8JH/\beta)}{\alpha B}},
\qquad \sum_{s\in\mathcal{I}} w_s = T.
\end{equation}
Together with equation~\eqref{eq:regret-probe-budget-bound}, the probe and radius contributions are at most $(15+6z)\,\overline{\Gamma}^{2}\,T\sqrt{JH\log(8JH/\beta)/(\alpha B)}$.

\paragraph{Changes between comparisons.}
It remains to bound the score-change contribution in
equation~\eqref{eq:regret-block-decomposition},
\begin{align}
\sum_{s\in\mathcal I}
\sum_{t=s}^{s+w_s-1}
\left[
\Psi_t(j_s)-\Psi_s(j_s)
+\Psi_s(j_s^\star)-\min_j\Psi_t(j)
\right].
\end{align}
This term tracks changes in both the retained candidate's score
and the best available score. The minimum score changes by at
most the largest change among individual candidate scores.
Hence, within a block,
\begin{align}
&
\Psi_t(j_s)-\Psi_s(j_s)
+\Psi_s(j_s^\star)-\min_j\Psi_t(j)
\notag\\
&\quad=
\Psi_t(j_s)-\Psi_s(j_s)
+\min_j\Psi_s(j)-\min_j\Psi_t(j)
\notag\\
&\quad\leq
|\Psi_t(j_s)-\Psi_s(j_s)|
+
\left|\min_j\Psi_s(j)-\min_j\Psi_t(j)\right|
\notag\\
&\quad\leq
2\sum_{\tau=s}^{t-1}
\max_j|\Psi_{\tau+1}(j)-\Psi_\tau(j)|
\qquad\text{(telescoping)}.
\end{align}
Summing over blocks gives
\begin{align}
&
\sum_{s\in\mathcal I}
\sum_{t=s}^{s+w_s-1}
\left[
\Psi_t(j_s)-\Psi_s(j_s)
+\Psi_s(j_s^\star)-\min_j\Psi_t(j)
\right]
\notag\\
&\quad\leq
2\sum_{s\in\mathcal I}
\sum_{\tau=s}^{s+w_s-2}
(s+w_s-1-\tau)
\max_j|\Psi_{\tau+1}(j)-\Psi_\tau(j)|
\notag\\
&\quad\leq
2(K-1)\sum_{t=0}^{T-2}
\max_j|\Psi_{t+1}(j)-\Psi_t(j)|.
\label{eq:regret-staleness-bound}
\end{align}
The first inequality interchanges the finite sums.
The second uses $w_s\leq K$ and the disjoint block interiors,
and empty sums are zero.

We bound the remaining score variation through parameter
movement and changes in the affordable shot allowance.
By Lemma~\ref{lem:risk-lipschitz} and
Assumption~\ref{ass:controller-updates},
\begin{align}
&
\sum_{t=0}^{T-2}
\max_j|\Psi_{t+1}(j)-\Psi_t(j)|
\notag\\
&\quad\leq
L_R\sum_{t=0}^{T-2}\|\theta_{t+1}-\theta_t\|_2
+
\lambda\overline\Gamma^2
\sum_{t=0}^{T-2}
\left|
N_{\mathrm{future},t+1}^{-1}
-N_{\mathrm{future},t}^{-1}
\right|
\notag\\
&\quad\leq
L_Rh\sum_{t=0}^{T-2}\frac1{t+1}
+
\lambda\overline\Gamma^2
\sum_{t=0}^{T-2}
\left|
N_{\mathrm{future},t+1}^{-1}
-N_{\mathrm{future},t}^{-1}
\right|.
\end{align}
The training allocation rule further implies
\begin{align}
N_{\mathrm{future},t+1}
&=
\frac{
(T-t)N_{\mathrm{future},t}
-\lfloor N_{\mathrm{future},t}\rfloor
}{T-t-1}
\notag\\
&=
N_{\mathrm{future},t}
+
\frac{
N_{\mathrm{future},t}
-\lfloor N_{\mathrm{future},t}\rfloor
}{T-t-1}
\geq N_{\mathrm{future},t}.
\end{align}
Thus,
\begin{align}
\sum_{t=0}^{T-2}
\left|
N_{\mathrm{future},t+1}^{-1}
-N_{\mathrm{future},t}^{-1}
\right|
&=
\frac1{N_{\mathrm{future},0}}
-\frac1{N_{\mathrm{future},T-1}}
&&\text{(monotonicity and telescoping)}
\notag\\
&\leq\frac1{N_{\mathrm{future},0}}.
\end{align}
Together with
$\sum_{t=0}^{T-2}(t+1)^{-1}\leq1+\log T$, this gives
\begin{align}
\sum_{t=0}^{T-2}
\max_j|\Psi_{t+1}(j)-\Psi_t(j)|
\leq
L_Rh(1+\log T)
+\frac{\lambda\overline\Gamma^2}{N_{\mathrm{future},0}}.
\label{eq:regret-variation-bound}
\end{align}
Substitution into equation~\eqref{eq:regret-staleness-bound}
therefore bounds the cumulative score-change contribution by
\begin{align}
2(K-1)
\left[
L_Rh(1+\log T)
+\frac{\lambda\overline\Gamma^2}{N_{\mathrm{future},0}}
\right].
\end{align}

\paragraph{Combining the bounds.}
The validation and probe events hold jointly with probability
at least $1-\beta$ by a union bound.
On this event, substituting the bounds on the validation, probe, radius and score-change into equation~\eqref{eq:regret-block-decomposition} yields
\begin{align}
\operatorname{Reg}_T
\leq{}&
(15+6z)\overline\Gamma^2T
\sqrt{\frac{JH\log(8JH/\beta)}{\alpha B}}
\notag\\
&+
2a_v(\beta)
\left(
\delta_\gamma\sum_{s\in\mathcal I}w_sD_s
+2T\varepsilon_{\mathrm{tol}}
\right)
\notag\\
&+
2(K-1)
\left[
L_Rh(1+\log T)
+\frac{\lambda\overline\Gamma^2}{N_{\mathrm{future},0}}
\right],
\end{align}
which proves the theorem.
\end{proof}

\restateresult{\corollaryrate*}

\begin{proof}[Proof of Corollary~\ref{cor:controller-regret-rate}]
Use the event of probability at least $1-\beta$ in
Theorem~\ref{thm:controller-regret}.
With $K=\lceil T^{2/3}\rceil$ and $T\geq2$,
\begin{align}
H=\left\lceil\frac TK\right\rceil
\leq\frac TK+1
\leq\frac{2T}{K}
\leq2T^{1/3}.
\end{align}
Since $B=\Theta(T)$, we write $B=\varrho T$ for a constant $\varrho>0$, and the probe term satisfies
\begin{align}
(15+6z)\,\overline\Gamma^2 T
\sqrt{\frac{JH\log(8JH/\beta)}{\alpha B}}
=
\mathcal O\!\left(
T^{2/3}\sqrt{\log(T/\beta)}
\right),
\end{align}
with the remaining parameters fixed.

For the validation term, $\kappa\in[0,1]$ implies
\begin{align}
D_s^2
={}&
\mathbb E_{X,X'\sim q_{\theta_s}^0}\kappa(X,X')
+\mathbb E_{Y,Y'\sim u}\kappa(Y,Y')
\notag\\
&-2\mathbb E_{X\sim q_{\theta_s}^0,Y\sim u}\kappa(X,Y)
\leq2.
\end{align}
Thus $\sum_s w_sD_s\leq\sqrt2\,T$, and
\begin{align}
&2a_v(\beta)
\left(\delta_\gamma\sum_s w_sD_s
+2T\varepsilon_{\mathrm{tol}}\right)
\notag\\
&\qquad\leq
2T a_v(\beta)
\left(\sqrt2\,\delta_\gamma+2\varepsilon_{\mathrm{tol}}\right)
=
\mathcal O\!\left(
T\sqrt{\frac{\log(2/\beta)}v}
\right).
\end{align}

Finally,
\begin{align}
N_{\mathrm{future},0}
=\frac{(1-\alpha)B}{P_{\mathrm{circ}}T}
=\frac{(1-\alpha)\varrho}{P_{\mathrm{circ}}},
\end{align}
which is independent of $T$.
Since $K-1<T^{2/3}$, the last term satisfies
\begin{align}
&2(K-1)
\left[
L_Rh(1+\log T)
+\frac{\lambda\overline\Gamma^2}{N_{\mathrm{future},0}}
\right]
\notag\\
&\qquad=
\mathcal O\!\left(T^{2/3}(1+\log T)\right).
\end{align}
Adding these three estimates yields
\begin{align}
\operatorname{Reg}_T
=
\mathcal O\!\left(
T^{2/3}
\bigl[1+\log T+\sqrt{\log(T/\beta)}\bigr]
+
T\sqrt{\frac{\log(2/\beta)}v}
\right),
\end{align}
which is
$\widetilde{\mathcal O}(T^{2/3}+T/\sqrt v)$.
\end{proof}

\end{document}